\documentclass[10pt]{iopart}

\usepackage{graphicx}
\usepackage{longtable}
\begin{document}

\title{Influence of the hexadecapole deformation on the two neutrino double-$\beta $
decay}
\author{R Chandra$^{1,2}$, P K Rath$^{2}$, P K Raina$^{3}$ and 
J G Hirsch$^{4}$}
\address{
$^{1}$Institute of Physics, Sachivalaya Marg, Bhubaneswar-751005, India\\
$^{2}$Department of Physics, University of Lucknow, Lucknow-226007, India\\
$^{3}$Department of Physics, IIT Kharagpur-721302, India\\
$^{4}$Instituto de Ciencias Nucleares, Universidad Nacional Aut\'{o}noma de
M\'{e}xico, A.P. 70-543 M\'{e}xico 04510 D.F., M\'{e}xico
}

\begin{abstract}
The two neutrino double beta $\left( \beta ^{-}\beta ^{-}\right) _{2\nu }$
decay of $\ ^{94,96}$Zr, $^{98,100}$Mo, $^{104}$Ru, $^{110}$Pd, $^{128,130}$%
Te and $^{150}$Nd nuclei for the $0^{+}\rightarrow 0^{+}$ transition is
studied in the PHFB model in conjunction with the pairing plus
quadrupole-quadrupole plus hexadecapole-hexadecapole effective two-body
interaction and the effect of the latter is investigated on the calculation
of nuclear transition matrix elements $M_{2\nu }$. The reliability of the
intrinsic wave functions of parent and daughter nuclei involved in the $%
\left( \beta ^{-}\beta ^{-}\right) _{2\nu }$ decay of\ above mentioned
nuclei is established by obtaining an overall agreement between a number of
theoretically calculated spectroscopic properties, namely the yrast spectra,
reduced $B(E2$:$0^{+}\rightarrow 2^{+})$ transition probabilities, static
quadrupole moments $Q(2^{+})$ and $g$-factors $g(2^{+})$ and the available
experimental data. The effect of deformation on $M_{2\nu }$ is also
investigated to inveterate its inverse relation with nuclear deformation.

\end{abstract}

\pacs {23.40.Hc, 21.60.Jz, 23.20.-g, 27.60.+j}
%\submitto{\JPG}
\maketitle
\section{Introduction}

\label{sec:level1}The nuclear $\beta \beta $ decay is a rare second order
semi-leptonic transition between two even $Z$-even $N$ isobars $_{Z}^{A}$X
and $_{Z\pm 2}^{~~A}$Y involving strangeness conserving charged weak
currents. The $\beta \beta $ decay can be broadly classified into four
experimentally distinguishable modes, namely two neutrino double beta $%
\left( \beta \beta \right) _{2\nu }$ decay \cite{maye35}, neutrinoless
double beta $\,\left( \beta \beta \right) _{0\nu }$ decay \cite{fury39},
single Majoron accompanied neutrinoless double beta $\,\left( \beta \beta
\phi \right) _{0\nu }$ decay \cite{chik80} and double Majoron accompanied
neutrinoless double beta $\left( \beta \beta \phi \phi \right) _{0\nu }$
decay \cite{aula82}. The$\left( \beta \beta \right) _{2\nu }$ decay
conserves the lepton number $L$ exactly and is an allowed process within the
standard model of electroweak unification (\textit{SM}). In $\left( \beta
\beta \right) _{0\nu }$ decay, the conservation of lepton number is violated
by two units and it is possible in models beyond the \textit{SM}, namely
GUTs (left-right symmetric SO(10), E(6) etc.), R$_{p}-$conserving as well as
violating SUSY models, leptoquark, compositeness and sterile neutrino
scenarios. In fact, the experimental observation of $\left( \beta \beta
\right) _{0\nu }$ decay would immediately imply that neutrinos are Majorana
particles and all the present experimental activities are directed towards
the observation of this particular decay mode. The $\,\left( \beta \beta
\phi \right) _{0\nu }$ and $\left( \beta \beta \phi \phi \right) _{0\nu }$
decay modes are possible processes in nine Majoron models as discussed by
Bamert and co-workers \cite{bame95}.

These decay modes can proceed via emission of two electrons ($\beta
^{-}\beta ^{-}$), emission of two positrons ($\beta ^{+}\beta ^{+}$),
electron-positron conversion ($\varepsilon \beta ^{+}$) and double electron
capture ($\varepsilon \varepsilon $). The latter three are energetically
competing modes and we refer to them as $e^{+}\beta \beta $ modes. There are
35 $\beta ^{-}\beta ^{-}$ and 34 $e^{+}\beta \beta $ emitters. Further, the
transitions in $\beta \beta $ decay modes may be from $0^{+}\rightarrow
J^{+} $ states. Presently, the study of $\beta ^{-}\beta ^{-}$ decay for the 
$0^{+}\rightarrow 0^{+}$ transition is the most preferable as the other
decay rates are suppressed due to kinematic reasons. However, the study of
other decay modes will be of importance in distinguishing the role of
different mechanisms involved in $\left( \beta \beta \right) _{0\nu }$ decay
once it is observed. Thus, the experimental as well as theoretical study of
nuclear $\beta \beta $ decay is quite wide in scope and has been excellently
reviewed over the past years [6-23]. In the present work, we restrict our
study to $\left( \beta ^{-}\beta ^{-}\right) _{2\nu }$ decay of $^{94,96}$%
Zr, $^{98,100}$Mo, $^{104}$Ru, $^{110}$Pd, $^{128,130}$Te and $^{150}$Nd
isotopes for the $0^{+}\rightarrow 0^{+}$ transition only.

The study of $\left( \beta ^{-}\beta ^{-}\right) _{2\nu }$ decay is quite
interesting from the nuclear structure point of view. The $\left( \beta
^{-}\beta ^{-}\right) _{2\nu }$ decay has been experimentally observed in
case of ten nuclei namely, $^{48}$Ca, $^{76}$Ge, $^{82}$Se, $^{96}$Zr, $%
^{100}$Mo, $^{116}$Cd, $^{128,130}$Te, $^{150}$Nd and $^{238}$U out of 35
possible candidates \cite{tret02} and one can extract nuclear transition
matrix elements (NTMEs) from the observed half-lives. Using the average
half-lives \cite{bara06} and the phase space factors \cite{boeh92}, the
extracted NTMEs $M_{2\nu }$ vary from 0.0152$\pm $0.0008 (0.0238$\pm $%
0.0013) to 0.1222$\pm $0.0034 (0.1909$\pm $0.0053) for $g_{A}=1.25$ (1.00)
corresponding to $^{130}$Te and $^{100}$Mo respectively. A comparison
between the theoretically calculated and experimentally extracted NTMEs
provides a cross-check on the reliability of different nuclear models used
for the calculation of NTMEs.

It is observed in all cases of $\left( \beta ^{-}\beta ^{-}\right) _{2\nu }$%
\ decay that the NTMEs$\ M_{2\nu }$ are sufficiently quenched. The
calculation of $M_{2\nu }$ requires the knowledge of the $\beta ^{-}$ or (p,
n) amplitude for the initial nucleus and the $\beta ^{+}$ or (n, p)
amplitude of the final nucleus, which in turn requires a complete set of
states of the intermediate nucleus in addition to the initial and final
nuclear states. The understanding as well as realization of this quenching
mechanism is the main motive of all the theoretical calculations. In solving
this problem, different nuclear models and nuclear structure scenarios have
been applied. Nuclear models, which are used in the calculation of NTMEs of $%
\left( \beta ^{-}\beta ^{-}\right) _{2\nu }$ decay, can be broadly
classified in to shell model and its variants, quasiparticle random phase
approximations (QRPA)\ and extensions to it and alternative models. The
details about these models -their advantages as well as limitations- have
been discussed by Suhonen and Civitarese \cite{suho98} and Faessler and
Simkovic \cite{faes98}.

The shell-model is the best choice for the calculation of the NTMEs as it
attempts to solve the nuclear many-body problem as exactly as possible. The
large scale shell model calculations by Caurier \textit{et al.} are more
realistic in which $^{76}$Ge, $^{82}$Se, $^{124}$Sn, $^{128}$Te, $^{130}$Te
and $^{136}$Xe have been studied \cite{caur99}. The $M_{2\nu }$ of $^{82}$Se
is calculated exactly and those of $^{76}$Ge and $^{136}$Xe are dealt in a
nearly exact manner \cite{caur96}. In the QRPA model, Vogel and Zirnbauer
were the first to provide an explanation of the observed suppression of $%
M_{2\nu }$ by a proper inclusion of ground state correlations through the
proton-neutron \textit{p-p} interaction in the \textit{S}=1, \textit{T}=0
channel and the calculated half-lives are in close agreement with all the
experimental data \cite{voge86}. The QRPA frequently overestimates the
ground state correlations and to cure the strong suppression of $M_{2\nu }$
due to increase in the attractive proton-neutron interaction, several
extensions of QRPA have been proposed. The most important proposals are
inclusion of proton-neutron pairing, renormalized QRPA, higher QRPA,
multiple commutator method (MCM) and particle number projection. However,
none of the above methods is free from ambiguities \cite{faes98}.
Alternative models, as the operator expansion method (OEM), the broken SU(4)
symmetry, two vacua RPA, the pseudo SU(3) and the single state dominance
hypothesis (SSDH) have their own problems \cite{suho98}.

The $\left( \beta ^{-}\beta ^{-}\right) _{0\nu }$ decay has not been
experimentally observed hitherto, and only limits on half-lives of $\left(
\beta ^{-}\beta ^{-}\right) _{0\nu }$ decay are available. Klapdor and his
collaborators have reported the $\left( \beta ^{-}\beta ^{-}\right) _{0\nu }$
decay of $^{76}$Ge in Heidelberg-Moscow experiment \cite{klap04}. However,
it is felt that the reported result needs independent verification by other
experiments \cite{aals02,elli04}. The observed half-life limits permit to
extract limits on various effective lepton number violating parameters,
namely Majorona neutrino mass, coupling of left and right handed weak
currents, mass of right handed heavy neutrino, mass of right handed W-boson,
intergeneration Yukawa coupling constants of SUSY models, leptoquark
coupling constants, compositeness scale, constraint on sterile neutrinos,
Majoron coupling constants, VEP and VLI parameters using the theoretically
calculated NTMEs of $\left( \beta ^{-}\beta ^{-}\right) _{0\nu }$ decay. In
order to get accurate effective lepton number violating parameters, one has
to calculate reliable NTMEs for $\left( \beta ^{-}\beta ^{-}\right) _{0\nu }$
decay. In the absence of experimental data, it is difficult to judge the
reliability of wave functions involved in the calculation of NTMEs of $%
\left( \beta ^{-}\beta ^{-}\right) _{0\nu }$ decay. Usually, the reliability
of wave functions is tested by reproducing the experimentally extracted
NTMEs $M_{2\nu }$ as both the modes involve same set of initial and final
wave functions although the nuclear transition operators are sensitive to
different spin-isospin correlations in $\left( \beta ^{-}\beta ^{-}\right)
_{2\nu }$ and $\left( \beta ^{-}\beta ^{-}\right) _{0\nu }$ decay modes.

The two main ingredients deciding the structure of nuclei participating in $%
\beta \beta $ decay are the pairing and deformation degrees of freedom. The
crucial role of deformation on NTMEs $M_{2\nu }$ has been predicted in the
case of $\left( \beta ^{-}\beta ^{-}\right) _{2\nu }$ decay of $^{100}$Mo
and $^{150}$Nd \cite{grif92,suho94}. The existence of an inverse correlation
between the GT strength and quadrupole moment has been already shown by
Auerbach \textit{et al} \cite{auer93} and Troltenier \textit{et al} \cite
{trol96}. The effect of deformation on the distribution of the Gamow-Teller
strength and $\beta $-decay properties has been studied using a
quasiparticle Tamm-Dancoff approximation (TDA) based on deformed
Hartree-Fock (DHF) calculations with Skyrme interactions \cite{fris95} and
in deformed self consistent HF+BCS+QRPA method with Skyrme type interactions 
\cite{sarr98}. N\'{a}cher \textit{et al} \cite{nach04} have presented a
novel method of deducing the deformation of $N=Z$ nucleus $^{76}$Sr, based
on the comparison of the experimental GT strength distribution $B(GT)$ from
its decay with the results of QRPA calculations. A deformed QRPA formalism,
using deformed Woods-Saxon potentials and deformed Skyrme Hartree-Fock mean
fields, was developed to describe simultaneously the energy distributions of
the single-$\beta $ GT strength and the $\left( \beta ^{-}\beta ^{-}\right)
_{2\nu }$ decay matrix elements \cite{alva04}. The deformation effect on the 
$\left( \beta ^{-}\beta ^{-}\right) _{2\nu }$ decay for ground-state
transition of $^{76}$Ge was studied in the framework of the deformed QRPA
with separable Gamow-Teller (GT) residual interaction \cite{pace04}.

In the light of above discussions, the Projected Hartree-Fock Bogoliubov
(PHFB) model is a convenient choice as an alternative model in which the
pairing and deformation degrees of freedom are incorporated on equal footing
and the rotational symmetry is restored by projection technique providing
wave functions with good angular momentum for the parent and daughter nuclei
involved in $\beta \beta $ decay. However, the PHFB model is unable to
provide information about the structure of the intermediate odd-odd nuclei
in its present version and hence, on the single $\beta $ decay rates and the
distribution of GT strength. In spite of this limitation, the PHFB model, in
conjunction with pairing plus quadrupole-quadrupole (\textit{PQQ})
interaction \cite{bara68}, has been successfully applied to study the $%
\left( \beta ^{\pm }\beta ^{\pm }\right) _{2\nu }$ decay for the $%
0^{+}\rightarrow 0^{+}$ transition where it was possible to describe the
lowest excited states of the parent and daughter nuclei along with their
electromagnetic transition strengths, as well as to reproduce their measured 
$\beta \beta $ decay rates \cite{chan05,rain06,sing07}. In the PHFB model,
the role of deformation in reproducing realistic NTMEs $M_{2\nu }$ has also
been investigated and it has been observed that there exists an inverse
correlation between the latter and the former \cite{chan05,rain06,sing07}.

In the present work, we add a hexadecapole-hexadecapole interaction term $%
V(HH)$ to the standard \textit{PQQ} interaction to check the stability of
our previous results of $\,\left( \beta ^{-}\beta ^{-}\right) _{2\nu }$
decay with respect to the change in effective two-body interaction. In
variation-after-projection (VAP) framework, the pairing plus
quadrupole-quadrupole plus hexadecapole-hexadecapole (\textit{PQQHH})
interaction has been successfully applied to study the yrast spectra of $%
^{68-76}$Ge, $^{72-78}$Se, $^{74-82}$Kr, $^{100-108}$Zr and $^{100-108}$Mo
isotopes \cite{hexa}. The present paper is organized as follows. In section
2, we present the theoretical formalism to calculate the NTME $M_{2\nu }$ in
the PHFB model in conjunction with summation method using \textit{PQQHH }%
interaction. The expressions to calculate the spectroscopic properties,
specifically, the yrast spectra, reduced $B(E2$:$0^{+}\rightarrow 2^{+})$
transition probabilities, static quadrupole moments $Q(2^{+})$ and $g$%
-factors $g(2^{+})$ are given by Dixit \textit{et al} \cite{dixi02}. The
calculated spectroscopic properties of $^{94,96}$Zr, $^{94,96,98,100}$Mo, $%
^{98,100,104}$Ru, $^{104,110}$Pd, $^{110}$Cd, $^{128,130}$Te, $^{128,130}$%
Xe, $^{150}$Nd and $^{150}$Sm nuclei are compared with the observed
experimental data in section 3 and there by, we check the ``goodness of wave
functions''. The same wave functions are used to study the $\left( \beta
^{-}\beta ^{-}\right) _{2\nu }$ decay of $^{94,96}$Zr, $^{98,100}$Mo, $%
^{104} $Ru, $^{110}$Pd, $^{128,130}$Te and $^{150}$Nd nuclei for the $%
0^{+}\rightarrow 0^{+}$ transition and the results are presented in the same
section. We also examine the effect of deformation on NTMEs $M_{2\nu }$ by
varying the strength of \textit{QQHH} part of the effective two-body
interaction in section 3. While presenting the theoretically calculated
results using \textit{PQQHH} interaction, we also give our previous results 
\cite{chan05,sing07} for comparison, which were calculated using \textit{PQQ}
interaction. Finally, we present some concluding remarks in section 4.

\section{Theoretical framework}

The inverse half-life of the $\left( \beta ^{-}\beta ^{-}\right) _{2\nu }$\
decay for the$\ 0^{+}\to 0^{+}$\ transition \cite{haxt84,doi85,tomo91} is
given by 
\begin{equation}
\lbrack T_{1/2}^{2\nu }(0^{+}\to 0^{+})]^{-1}=G_{2\nu }|M_{2\nu }|^{2}
\end{equation}
where the phase space factor $G_{2\nu }$\ can be calculated with good
accuracy \cite{doi85,boeh92} and the nuclear model dependent NTME $M_{2\nu }$
is written as 
\begin{eqnarray}
M_{2\nu } &=&\sum\limits_{N}\frac{\langle 0_{F}^{+}||\mathbf{\sigma}\tau
^{+}||1_{N}^{+}\rangle \langle 1_{N}^{+}||\mathbf{\sigma }\tau
^{+}||0_{I}^{+}\rangle }{E_{N}-(E_{I}+E_{F})/2}  \nonumber \\
&=&\sum\limits_{N}\frac{\langle 0_{F}^{+}||\mathbf{\sigma }\tau
^{+}||1_{N}^{+}\rangle \langle 1_{N}^{+}||\mathbf{\sigma }\tau
^{+}||0_{I}^{+}\rangle }{E_{0}+E_{N}-E_{I}}  \label{m2n}
\end{eqnarray}
with $\ $%
\begin{equation}
E_{0}=\frac{1}{2}\left( E_{I}-E_{F}\right) =\frac{1}{2}Q_{\beta \beta }+m_{e}
\end{equation}
To evaluate (\ref{m2n}), it is required to sum over all the 
intermediate 
$1_{N}^{+}$ states . However, it is not possible to study the structure of
odd-odd nuclei in the present version of the PHFB model. Hence, we carry out
the summation over the intermediate states by using the summation method
given by Civitarese and Suhonen\textit{\ }\cite{civi93}. Using summation
method, the $M_{2\nu }$ is expressed as 
\begin{equation}
M_{2\nu }=\frac{1}{E_{0}}\left\langle 0_{F}^{+}\left| \sum_{\mu }(-1)^{\mu
}\Gamma _{-\mu }F_{\mu }\right| 0_{I}^{+}\right\rangle  \label{m2nsu}
\end{equation}
where $\Gamma _{\mu }$ is given by 
\begin{equation}
\Gamma _{\mu }=\sigma _{\mu }\tau ^{+}
\end{equation}
and 
\begin{equation}
F_{\mu }=\sum_{\lambda =0}^{\infty }\frac{(-1)^{\lambda }}{E_{0}^{\lambda }}%
D_{\lambda }\Gamma _{\mu }
\end{equation}
with 
\begin{equation}
D_{\lambda }\Gamma _{\mu }=\left[ H,\left[ H,........,\left[ H,\Gamma _{\mu
}\right] .......\right] \right] ^{(\lambda \hbox{ times})}
\end{equation}
When the GT operator commutes with the effective two-body interaction, the
(\ref{m2nsu}) can be further simplified to 
\begin{equation}
M_{2\nu }=\sum\limits_{\pi ,\nu }\frac{\langle 0_{F}^{+}||\mathbf{\sigma
\cdot \sigma }\tau ^{+}\tau ^{+}||0_{I}^{+}\rangle }{E_{0}+\varepsilon
(n_{\pi },l_{\pi },j_{\pi })-\varepsilon (n_{\nu },l_{\nu },j_{\nu })}
\label{de}
\end{equation}

\noindent In the case of pseudo-SU(3) model, the GT operator commutes with
the two-body interaction \cite{cast94,hirs95,cero99} and the energy
denominator is a well-defined quantity without any free parameter. It has
been evaluated exactly for $\left( \beta ^{-}\beta ^{-}\right) _{2\nu }$ 
\cite{cast94,hirs95} and $\left( \varepsilon \varepsilon \right) _{2\nu }$
modes \cite{cero99} in the pseudo-SU(3) scheme.

In the present work, we use a Hamiltonian with \textit{PQQHH} type of
effective two-body interaction. The Hamiltonian is explicitly written as 
\begin{equation}
H=H_{sp}+V(P)+\zeta _{qq}\left[ V(QQ)+V(HH)\right]  \label{hmtn}
\end{equation}
\smallskip \noindent where$\ H_{sp}$ denotes the single particle Hamiltonian$%
.$ The pairing part of the effective two-body interaction $V(P)$ is given by 
\begin{equation}
V{(}P{)}=-\left( \frac{G}{4}\right) \sum\limits_{\alpha \beta
}(-1)^{j_{\alpha }+j_{\beta }-m_{\alpha }-m_{\beta }}a_{\alpha }^{\dagger
}a_{\bar{\alpha}}^{\dagger }a_{\bar{\beta}}a_{\beta }  \label{h}
\end{equation}

\noindent where $\alpha $ denotes the quantum numbers ($nljm$) and the state 
$\bar{\alpha}$ is same as $\alpha $ but with the sign of $m$ reversed. The 
\textit{QQ} part of the effective interaction $V(QQ)$\ is expressed as 
\begin{equation}
V(QQ)=-\left( \frac{\chi _{2}}{2}\right) \sum\limits_{\alpha \beta \gamma
\delta }\sum\limits_{\mu }(-1)^{\mu }\langle \alpha |q_{2\mu }|\gamma
\rangle \langle \beta |q_{2-\mu }|\delta \rangle \ a_{\alpha }^{\dagger
}a_{\beta }^{\dagger }\ a_{\delta }\ a_{\gamma }  \label{i}
\end{equation}

\noindent where 
\begin{equation}
q{_{2\mu }}=\left( \frac{16\pi }{5}\right) ^{1/2}r^{2}Y_{2\mu }(\theta ,\phi
)
\end{equation}
The \textit{HH} part of the effective interaction $V(HH)$ is given as 
\begin{equation}
V(HH)=-\left( \frac{\chi _{4}}{2}\right) \sum\limits_{\alpha \beta \gamma
\delta }\sum\limits_{\nu }(-1)^{\nu }\langle \alpha |q_{4\nu }|\gamma
\rangle \langle \beta |q_{4-\nu }|\delta \rangle \ a_{\alpha }^{\dagger
}a_{\beta }^{\dagger }\ a_{\delta }\ a_{\gamma }
\end{equation}

\noindent with 
\begin{equation}
q{_{4\nu }}=r^{4}Y_{4\nu }(\theta ,\phi )
\end{equation}
Further, $\zeta _{qq}$ is an arbitrary parameter and the final results are
obtained by setting $\zeta _{qq}=1$. The purpose of introducing it is to
study the effect of deformation by varying the strength of effective
two-body \textit{QQHH }interaction.

The model Hamiltonian used in the present work does not commute with the GT
operator. Hence, the energy denominator is not a well-defined quantity.
However, the violation of isospin symmetry for the \textit{QQHH} part of our
model Hamiltonian is negligible as will be evident from the parameters of
the two-body interaction given in section 3. Further, the violation of
isospin symmetry for the pairing part of the two-body interaction is
presumably small. With these assumptions, the NTME $M_{2\nu }$ of $\left(
\beta ^{-}\beta ^{-}\right) _{2\nu }$ decay for the 0$^{+}\rightarrow 0^{+}$
transition in the PHFB model in conjunction with the summation method can be
obtained as follows.

In the PHFB model, states with good angular momentum $\mathbf{J}$ are
obtained from the axially symmetric HFB intrinsic state ${|\Phi _{0}\rangle }
$ with \textit{K}=0 using the standard projection technique \cite{onis66}
given by 
\begin{equation}
\left| \Psi _{00}\right\rangle =\frac{2J+1}{8\pi ^{2}}\int D_{00}^{J}(\Omega
)R(\Omega )\left| \Phi _{0}\right\rangle d\Omega
\end{equation}
where $R(\Omega )$\ and $\ D_{00}^{J}(\Omega )$\ are the rotation operator
and the rotation matrix respectively. The axially symmetric HFB intrinsic
state ${|\Phi _{0}\rangle }$ can be written as 
\begin{equation}
{|\Phi _{0}\rangle }=\prod\limits_{im}(u_{im}+v_{im}b_{im}^{\dagger }b_{i%
\bar{m}}^{\dagger })|0\rangle
\end{equation}
where the creation operators $\ b_{im}^{\dagger }$\ and $\ b_{i\bar{m}%
}^{\dagger }$\ are defined as 
\begin{equation}
b{_{im}^{\dagger }}=\sum\limits_{\alpha }C_{i\alpha ,m}a_{\alpha
,m}^{\dagger }\quad \hbox{and}\mathrm{\quad }b_{i\bar{m}}^{\dagger
}=\sum\limits_{\alpha }(-1)^{l+j-m}C_{i\alpha ,m}a_{\alpha ,-m}^{\dagger }
\end{equation}

\noindent Finally, one obtains the expression for the NTME $M_{2\nu }$ of $%
\left( \beta ^{-}\beta ^{-}\right) _{2\nu }$ decay for the $0^{+}\rightarrow
0^{+}$ transition as \cite{chan05,sing07} 
\begin{eqnarray}
M_{2\nu } &=&\sum\limits_{\pi ,\nu }\frac{\langle {\Psi _{00}^{J_{f}=0}}||%
\mathbf{\sigma \cdot \sigma }\tau ^{+}\tau ^{+}||{\Psi _{00}^{J_{i}=0}}%
\rangle }{E_{0}+\varepsilon (n_{\pi },l_{\pi },j_{\pi })-\varepsilon (n_{\nu
},l_{\nu },j_{\nu })}  \nonumber \\
&=&\left[ n_{(Z,N)}^{Ji=0}n_{(Z+2,N-2)}^{J_{f}=0}\right]
^{-1/2}\int\limits_{0}^{\pi }n_{(Z,N),(Z+2,N-2)}(\theta )\sum\limits_{\alpha
\beta \gamma \delta }\frac{\left\langle \alpha \beta \left| \mathbf{\sigma }%
_{1}.\mathbf{\sigma }_{2}\tau ^{+}\tau ^{+}\right| \gamma \delta
\right\rangle }{E_{0}+\varepsilon _{\alpha }(n_{\pi },l_{\pi },j_{\pi
})-\varepsilon _{\gamma }(n_{\nu },l_{\nu },j_{\nu })}  \nonumber \\
&&\times \sum\limits_{\varepsilon \eta }\frac{\left( f_{Z+2,N-2}^{(\pi
)*}\right) _{\varepsilon \beta }\left( F_{Z,N}^{(\nu )*}\right) _{\eta
\delta }}{\left[ \left( 1+F_{Z,N}^{(\pi )}(\theta )f_{Z+2,N-2}^{(\pi
)*}\right) \right] _{\varepsilon \alpha }\left[ \left( 1+F_{Z,N}^{(\nu
)}(\theta )f_{Z+2,N-2}^{(\nu )*}\right) \right] _{\gamma \eta }}sin\theta
d\theta  \label{m2nhf}
\end{eqnarray}
where 
\begin{equation}
n^{J}=\int\limits_{0}^{\pi }\left[ det\left( 1+F^{(\pi )}f^{(\pi )^{\dagger
}}\right) \right] ^{1/2}\left[ det\left( 1+F^{(\nu )}f^{(\nu )^{\dagger
}}\right) \right] ^{1/2}d_{00}^{J}(\theta )sin(\theta )d\theta
\end{equation}
and 
\begin{equation}
n_{(Z,N),(Z+2,N-2)}{(\theta )}=\left[ det\left( 1+F_{Z,N}^{(\nu
)}f_{Z+2,N-2}^{(\nu )^{\dagger }}\right) \right] ^{1/2}\times \left[
det\left( 1+F_{Z,N}^{(\pi )}f_{Z+2,N-2}^{(\pi )^{\dagger }}\right) \right]
^{1/2}
\end{equation}
The $\pi (\nu )$\ represents the proton (neutron) of nuclei involved in the $%
\left( \beta ^{-}\beta ^{-}\right) _{2\nu }$ decay process. The matrices $%
f_{Z,N}$ and $F_{Z,N}(\theta )$ are given by 
\begin{eqnarray}
f_{Z,N} &=&\sum\limits_{i}C_{ij_{\alpha },m_{\alpha }}C_{ij_{\beta
},m_{\beta }}\left( v_{im_{\alpha }}/u_{im_{\alpha }}\right) \delta
_{m_{\alpha },-m_{\beta }}  \label{ff} \\
F_{Z,N}(\theta ) &=&\sum\limits_{m_{\alpha }^{\prime }m_{\beta }^{\prime
}}d_{m_{\alpha },m_{\alpha }^{\prime }}^{j_{\alpha }}(\theta )d_{m_{\beta
},m_{\beta }^{\prime }}^{j_{\beta }}(\theta )f_{j_{\alpha }m_{\alpha
}^{\prime },j_{\beta }m_{\beta }^{\prime }}  \label{fff}
\end{eqnarray}

With the assumption that the difference in single particle energies of
protons in the intermediate nucleus and neutrons in the parent nucleus is
mainly due to the difference in Coulomb energies, one obtains 
\begin{equation}
\varepsilon (n_{\pi },l_{\pi },j_{\pi })-\varepsilon (n_{\nu },l_{\nu
},j_{\nu })=\left\{ 
\begin{array}{llll}
\Delta _{C} &  & for & n_{\nu }=n_{\pi },l_{\nu }=l_{\pi },j_{\nu }=j_{\pi }
\\ 
\Delta _{C}+\Delta E_{s.o.splitting} &  & for & n_{\nu }=n_{\pi },l_{\nu
}=l_{\pi },j_{\nu }\neq j_{\pi } \label{den}
\end{array}
\right. ,
\end{equation}
where the Coulomb energy difference $\Delta _{C}$ is given by Bohr and
Mottelson as \cite{bohr98}. 
\begin{equation}
\Delta _{C}=\frac{0.70}{A^{1/3}}\left[ \left( 2Z+1\right) -0.76\left\{
\left( Z+1\right) ^{4/3}-Z^{4/3}\right\} \right] MeV
\end{equation}

The numerical calculation of $M_{2\nu }$ for the nuclei involved in the $%
\left( \beta ^{-}\beta ^{-}\right) _{2\nu }$ decay involves the setup of
matrices $f_{Z,N}$ and $F_{Z,N}(\theta )$ given by (\ref{ff}) 
and (\ref {fff}) at 20 Gaussian quadrature points in the range 
($0$, $\pi $) using the results of PHFB calculations, which are 
summarized by amplitudes $(u_{im},v_{im})$\ and expansion coefficients 
$C_{ij,m}$. Subsequently, the required NTME is evaluated in a 
straightforward manner using (\ref{m2nhf}).

It must be underlined that, in the present context, the use of the summation
method goes beyond the closure approximation, because each proton-neutron
excitation is weighted depending on its spin-flip or non-spin-flip
character. The explicit inclusion of the spin-orbit splitting in the energy
denominator, (\ref{den}), implies that it cannot be factorized out of
the sum in (\ref{m2n}). In this sense, employing the summation 
method in
conjunction with the PHFB formalism is richer than what was done in previous
application with the pseudo SU(3) model \cite{cast94,hirs95}.

\section{Results and discussions}

The model space and single particle energies (SPE's) are the same as in our
earlier calculation on $\left( \beta ^{-}\beta ^{-}\right) _{2\nu }$ decay
for the $0^{+}\rightarrow 0^{+}$ transition \cite{chan05,sing07}. However,
we briefly discuss in the following the model space and single particle
energies (SPE's) used to generate the HFB wave functions for convenience. We
treat the doubly even nucleus $^{76}$Sr ($N=Z=38$) as an inert core in case
of $^{94,96}$Zr, $^{94,96,98,100}$Mo, $^{98,100,104}$Ru, $^{104,110}$Pd and $%
^{110}$Cd nuclei, with the valence space spanned by $1p_{1/2},$ $2s_{1/2,}$ $%
1d_{3/2},$ $1d_{5/2},$ $0g_{7/2},$ $0g_{9/2}$ and $0h_{11/2}$ orbits for
protons and neutrons. The $1p_{1/2}$ orbit has been included in the valence
space to examine the role of the $Z=40$ proton core vis-a-vis the onset of
deformation in the highly neutron rich isotopes. The set of single particle
energies (SPE's) used here are (in MeV) $\varepsilon (1p_{1/2})=-0.8$, $%
\varepsilon (0g_{9/2})=0.0$, $\varepsilon (1d_{5/2})=5.4$, $\varepsilon
(2s_{1/2})=6.4$, $\varepsilon (1d_{3/2})=7.9$, $\varepsilon (0g_{7/2})=8.4$
and $\varepsilon (0h_{11/2})=8.6$ for proton and neutron.

In case of $^{128,130}$Te, $^{128,130}$Xe, $^{150}$Nd and $^{150}$Sm nuclei,
we treat the doubly even nucleus $^{100}$Sn ($N=Z=50$) as an inert core with
the valence space spanned by $2s_{1/2}$, $1d_{3/2}$, $1d_{5/2}$, $1f_{7/2},$ 
$0g_{7/2}$, $0h_{9/2}$ and $0h_{11/2}$ orbits for protons and neutrons. The
change of model space is forced upon as in the model space used for mass
region $A\approx 100$, the number of neutrons increase to about 40 for
nuclei occurring in the mass region $A\approx 130$. With the increase in
neutron number, the yrast energy spectra gets compressed due to increase in
the attractive part of effective two-body interaction. The set of single
particle energies (SPE's) used here are in MeV: $\varepsilon (1d_{5/2})=0.0$%
, $\varepsilon (2s_{1/2})=1.4$, $\varepsilon (1d_{3/2})=2.0$, $\varepsilon
(0g_{7/2})=4.0$, $\varepsilon (0h_{11/2})=6.5$ (4.8 for $^{150}$Nd and $%
^{150}$Sm), $\varepsilon (1f_{7/2})=12.0$ (11.5 for $^{150}$Nd and $^{150}$%
Sm), $\varepsilon (0h_{9/2})=12.5$ (12.0 for $^{150}$Nd and $^{150}$Sm) for
proton and neutron.

The HFB\ wave functions are generated using an effective Hamiltonian with 
\textit{PQQHH} type of two-body interaction. Explicitly, the Hamiltonian can
be written as 
\begin{equation}
H=H_{sp}+V(P)+\zeta _{qq}\left[ V(QQ)+V(HH)\right] 
\end{equation}
where $H_{sp}$ denotes the single particle Hamiltonian. The $V(P)$, $V(QQ)$
and $V(HH)$ represent the pairing, quadrupole-quadrupole and
hexadecapole-hexadecapole part of the effective two-body interaction. The $\
\zeta _{qq}$ is an arbitrary parameter and the final results are obtained by
setting the$\ \zeta _{qq}=1$. The purpose of introducing $\zeta _{qq}$ is to
study the role of deformation by varying the strength of \textit{QQ}
interaction. The strengths of the pairing interaction is fixed through the
relation $G_{p}=30/A$ MeV and $G_{n}=20/A$ MeV, which are same as used by
Heestand \textit{et al.} \cite{hees69} to explain the experimental $g(2^{+})$
data of some even-even Ge, Se, Mo, Ru, Pd, Cd and Te isotopes in Greiner's
collective model \cite{grei66}. For $^{94}$Zr and $^{96}$Zr, we have used $%
G_{n}=18/A$ and $22/A$ MeV respectively. The strengths of the pairing
interaction fixed for $^{128,130}$Te, $^{128,130}$Xe, $^{150}$Nd and $^{150}$%
Sm are $G_{p}=35/A$ MeV and $G_{n}=35/A$ MeV.

The strengths of the like particle components of the \textit{QQ} interaction
are taken as $\chi _{2pp}=\chi _{2nn}=0.0105$ MeV \textit{b}$^{-4}$, where 
\textit{b} is oscillator parameter. The strength of proton-neutron (\textit{%
pn}) component of the \textit{QQ} interaction $\chi _{2pn}$ is varied so as
to obtain the spectra of considered nuclei namely $^{94,96}$Zr, $%
^{94,96,98,100}$Mo, $^{98,100,104}$Ru, $^{104,110}$Pd, $^{110}$Cd, $%
^{128,130}$Te, $^{128,130}$Xe, $^{150}$Nd and $^{150}$Sm in optimum
agreement with the experimental results. The theoretical spectra has been
taken to be the optimum one if the excitation energy of the $\ $2$^{+}$
state \ $E_{2^{+}}$ is reproduced as closely as possible to the experimental
value. The prescribed set of parameters for the strength of \textit{QQ}
interaction are consistent with those of Arima suggested on the basis of an
empirical analysis of effective two-body interaction \cite{arim81}.

The relative magnitudes of the parameters of the \textit{HH} part of the two
body interaction are calculated from a relation suggested by Bohr and
Mottelson \cite{bohr75}. According to them the approximate magnitude of
these constants for isospin $T=0$ is given by 
\begin{equation}
\chi _{\lambda }=\frac{4\pi }{2\lambda +1}\frac{m\omega _{0}^{2}}{%
A\left\langle r^{2\lambda -2}\right\rangle }\,\,\,\,\,\,\,\,\,\mathrm{{%
for\,\,}\lambda =1,2,3,4\cdot \cdot \cdot }
\end{equation}
and the parameters for the $T=1$ case are approximately half of their $T=0$
counterparts. We take the $\chi _{4}$ for $T=1$ case as exactly half of the $%
T=0$ case. Using $b=1.0032A^{1/6}$, one obtains 
\begin{eqnarray}
\chi _{4} &=&\left[ \left( \frac{16}{25}\right) \left( \frac{2}{3}\right)
^{2/3}\right] \chi _{2}A^{-2/3}b^{-4}  \nonumber \\
&=&0.4884\chi _{2}A^{-2/3}b^{-4}
\end{eqnarray}
We fix $\chi _{2pn}$ through the experimentally available energy spectra for
a given model space, SPE's, $G_{p}$, $G_{n}$ and $\chi _{2pp}$. We present
the values of $\chi _{2pn}$ in table 1. All these input parameters are kept
fixed to calculate other spectroscopic properties. Further, we have
performed independent calculations for the parent and daughter nuclei
involved in the $\beta \beta $ decay, whose deformations are in general
different.

\subsection{Yrast spectra and electromagnetic properties}

The calculated values of excitation energies $E_{2^{+}}$, $E_{4^{+}}$ and $%
E_{6^{+}}$ for all the nuclei of interest along with the experimental ones 
\cite{saka84} are given in table 1. It can be seen from table that the
theoretical spectra is more expanded in comparison to the experimental
spectra for all nuclei although the agreement between the theoretically
calculated and experimentally observed $E_{2^{+}}$ is quite good. This can
be taken care in conjunction with the VAP prescription \cite{vap}. However,
our aim is to reproduce the properties of only low-lying 2$^{+}$ state and
hence, we do not invoke the VAP prescription, which will unnecessarily
complicate the numerical calculation.

The reduced transition probabilities $B(E2$:$0^{+}\rightarrow 2^{+})$ are
calculated for effective charges $e_{eff}=$ 0.40, 0.50 and 0.60. The
experimentally observed results \cite{rama87,rama01} are also given in the
same table. It is noticed that the calculated and the observed $B(E2$:$%
0^{+}\to 2^{+})$ values are in excellent agreement in case of $^{94}$Zr, $%
^{94}$Mo, $^{104}$Ru and $^{104}$Pd isotopes for $\ e_{eff}=0.60$. For the
same $\ e_{eff}$, the theoretically calculated $B(E2$:$0^{+}\rightarrow
2^{+})$ differ by 0.007 and 0.004 e$^{2}$b$^{2}$ only from the experimental
data for $^{100}$Mo and $^{100}$Ru isotopes. In case of $^{96}$Zr, $^{96}$%
Mo, $^{110}$Cd, $^{128,130}$Te, $^{128,130}$Xe and $^{150}$Nd isotopes, the
calculated $B(E2$:$0^{+}\to 2^{+})$ agree with experimentally observed
values at $e_{eff}=0.50$. However, the calculated $B(E2$:$0^{+}\to 2^{+})$
differ by 0.056 and 0.508 e$^{2}$b$^{2}$ from the experimental results in
case of $^{110}$Pd and $^{150}$Sm nuclei for $e_{eff}=0.50$. The calculated $%
B(E2$:$0^{+}\to 2^{+})$ of $^{98}$Mo and $^{98}$Ru nuclei for $e_{eff}=0.40$
are in agreement with the experimental values.

The theoretically calculated $\ Q(2^{+})$, which are calculated for the same
effective charges, i.e. $e_{eff}=$ 0.40, 0.50 and 0.60, and the experimental 
$\ Q(2^{+})$ results \cite{ragh89} are given in the same table 2. No
experimental $Q(2^{+})$ result is available for $^{94,96}$Zr and $^{128,130}$%
Xe nuclei. For the same effective charge as used in case of $B(E2$:$0^{+}\to
2^{+})$, the agreement between the calculated and experimental $Q(2^{+})$ is
good for $^{104}$Ru, $^{110}$Pd and $^{150}$Sm nuclei. The calculated $%
Q(2^{+})$ are off by 0.10, 0.135, 0.02 and 0.021 eb in case of $^{98,100}$%
Mo, $^{100}$Ru and $^{150}$Nd nuclei respectively from the experimental
values. The theoretical $Q(2^{+})$ results are quite off from the observed
values for the rest of nuclei.

The gyromagnetic factors $g(2^{+})$ are calculated with $g_{l}^{\pi }=1.0$, $%
g_{l}^{\nu }=0.0$ and $g_{s}^{\pi }=g_{s}^{\nu }=0.60$ and presented in
table 2 along with the available experimental $g(2^{+})$ data \cite{ragh89}.
No experimental result is available for $^{96}$Zr and $^{94,96}$Mo. The
theoretical $g(2^{+})$ value of $^{94}$Zr is a pathological case. The
calculated $g(2^{+})$ is 0.112 nm while the most recent measured value is $%
-0.329\pm 0.015$ nm \cite{spei02}. The calculated and experimentally observed%
$\ g(2^{+})$ results are in good agreement for $^{98,100}$Mo, $^{98}$Ru, $%
^{104}$Pd, $^{110}$Cd and $^{128}$Xe nuclei. The discrepancy between
calculated and experimentally observed $\ g(2^{+})$ values are 0.047, 0.015,
0.089 and 0.028 nm only for $^{100,104}$Ru, $^{110}$Pd and $^{130}$Xe nuclei
respectively. The theoretically calculated and experimentally observed $\
g(2^{+})$ results are off by 0.136, 0.257, 0.161 and 0.161 nm in case of $%
^{128,130}$Te, $^{150}$Nd and $^{150}$Sm nuclei respectively.

It is observed from table 1 and table 2 that the calculated yrast spectra,
reduced $B(E2$:$0^{+}\to 2^{+})$ transition probabilities, static quadrupole
moments $Q(2^{+})$ and gyromagnetic factors $g(2^{+})$ with \textit{PQQ} and 
\textit{PQQHH }type of effective two body interaction do not differ much.
This establishes the stability of our previous results \cite{chan05,sing07}
against the change in effective two-body interaction.

\subsection{Results of $\ $ $\left( \beta ^{-}\beta ^{-}\right) _{2\nu }$
decay}

For the calculation of $\left( \beta ^{-}\beta ^{-}\right) _{2\nu }$ decay
rates, the phase space factors $G_{2\nu }$ of $^{94,96}$Zr, $^{98,100}$Mo, $%
^{104}$Ru, $^{110}$Pd, $^{128,130}$Te and $^{150}$Nd nuclei for the 0$%
^{+}\rightarrow $0$^{+}$ transition are 2.304$\times 10^{-21}$ yr$^{-1}$,
1.927$\times 10^{-17}$ yr$^{-1}$, 9.709$\times 10^{-29}$ yr$^{-1}$, 9.434$%
\times $10$^{-18}$ yr$^{-1}$, 9.174$\times 10^{-21}$ yr$^{-1}$, 3.984$\times
10^{-19}$ yr$^{-1}$, 8.475$\times 10^{-22}$ yr$^{-1}$, 4.808$\times 10^{-18}$
yr$^{-1}$ and 1.189$\times 10^{-16}$ yr$^{-1}$ respectively for $\ g_{A}$=
1.25 \cite{boeh92}. However, it is more justified to use the nuclear matter
value of $\ g_{A}$ around 1.0 in heavy nuclei. Hence, the experimental $%
M_{2\nu }$ as well as the theoretical $T_{1/2}^{2\nu }$ are calculated for $%
\ g_{A}=1.0$ and 1.25. In table 3, we compile all the available experimental
and the theoretical results along with our calculated $M_{2\nu }$ and
corresponding half-lives $\ T_{1/2}^{2\nu }$ of all the nuclei under
consideration for the $0^{+}\to 0^{+}$ transition. We also present the $%
M_{2\nu }$ extracted from the experimentally observed $T_{1/2}^{2\nu }$
using the given phase space factors. We present only the theoretical $%
T_{1/2}^{2\nu }$ for those models for which no direct or indirect
information about $M_{2\nu }$ is available to us.

The $\left( \beta ^{-}\beta ^{-}\right) _{2\nu }$ decay of $^{94}$Zr for the$%
\ 0^{+}\to 0^{+}$ transition has been investigated experimentally only by
Arnold \textit{et al} \cite{arno99} and the reported limit is $T_{1/2}^{2\nu
}>1.1\times 10^{17}$ yr. The half-life calculated in the PHFB model for $%
g_{A}=1.25$ is $1.086\times 10^{23}$ yr, which lies within the range given
by Bobyk \textit{et al }\cite{boby00}. The calculated half-life $%
T_{1/2}^{2\nu }$ in QRPA model \cite{stau90} is smaller than the presently
calculated $T_{1/2}^{2\nu }$ by a factor of 1.5 approximately for $%
g_{A}=1.25 $. On the other hand, the half-life $T_{1/2}^{2\nu }$ calculated
in OEM \cite{hirs94} is larger than our PHFB model value for $g_{A}=1.25$ by
a factor of 15 approximately. The predicted $T_{1/2}^{2\nu }$ for $g_{A}=1.0$
in the PHFB model is $2.652\times 10^{23}$ yr .

In the case of $^{96}$Zr$\rightarrow ^{96}$Mo, the theoretically calculated $%
M_{2\nu }$ given by Stoica using SRPA \cite{stoi95} is too small in
comparison to the NTME extracted from the experimental data. On the other
hand, the calculated half-life $T_{1/2}^{2\nu }$ in the OEM \cite{hirs94} is
quite off from the observed experimentally observed result. The $M_{2\nu }$
calculated by Engel \textit{et al} \cite{enge88} and Barabash \textit{et al }%
\cite{bara96} using QRPA are close to the experimentally observed lower
limit of Wieser \textit{et al }\cite{wies01} for $g_{A}=1.0$. The $%
T_{1/2}^{2\nu }$ calculated by in RQRPA using WS basis and AWS basis \cite
{toiv97} are $4.2\times 10^{19}$ yr and $4.4\times 10^{19}$ yr respectively
and agree with the experimental $T_{1/2}^{2\nu }$ of Kawashima \textit{et al}
\cite{kawa93}. The $T_{1/2}^{2\nu }$ predicted by Staudt \textit{et al} \cite
{stau90} is in agreement with the experimental results of Barabash \cite
{bara98} and Wieser \textit{et al }\cite{wies01}. The experimentally
observed $T_{1/2}^{2\nu }$ given by NEMO 3 \cite{lala05} and Wieser \textit{%
et al }\cite{wies01} are favored by the $T_{1/2}^{2\nu }$ calculated in the
PHFB model and SU(4)$_{\sigma \tau }$ \cite{rumy98} for $g_{A}=1.25$. The
predicted half-life $T_{1/2}^{2\nu }$ of Bobyk \textit{et al }\cite{boby00}
has a wide range and cover all the available experimental results.

No experimental result for $T_{1/2}^{2\nu }$ of $^{98}$Mo isotope is
available so far. It has been studied theoretically in the QRPA \cite{stau90}%
, OEM \cite{hirs94} and SRQRPA \cite{boby00}. The presently calculated $%
T_{1/2}^{2\nu }$ $=$ 6.36$\times $10$^{29}-$1.552$\times $10$^{30}$ yr for $%
g_{A}=1.25-1.0$ respectively is within the range given by SRQRPA \textit{\ }%
model. The calculated $T_{1/2}^{2\nu }$ in QRPA and OEM are larger than our
predicted value for $g_{A}=1.0$ by approximately a factor of 2 and 4
respectively.

In the case of $^{100}$Mo, the theoretically calculated $M_{2\nu }$ in SRPA 
\cite{stoi95} is too small in comparison to the experimental $M_{2\nu }$.
The $\left( \beta ^{-}\beta ^{-}\right) _{2\nu }$ decay rate of $^{100}$Mo
calculated by Staudt \textit{et al }in QRPA \cite{stau90}\textit{\ }and
Hirsch \textit{et al} in OEM \cite{hirs94} are off from the experimental $%
T_{1/2}^{2\nu }$. The calculated $M_{2\nu }$ by Griffiths \textit{et al }%
\cite{grif92}\textit{\ }using QRPA model favors the results of INS Baksan 
\cite{vasi90} and LBL \cite{garc93} for $g_{A}=1.0$. On the other hand, the $%
M_{2\nu }$ predicted by Engel \textit{et al} \cite{enge88} and Civitarese 
\textit{et al} \cite{civi98} are in agreement with the results of LBL \cite
{garc93}, LBL collaboration \cite{alst97}, UC Irvine \cite{silv97} and
ITEP+INFN \cite{ashi01} for $g_{A}=1.0$. The NTMEs $M_{2\nu }$ predicted in
SU(4)$_{\sigma \tau }$ \cite{rumy98} and pseudo-SU3 using spherical
occupation wave functions \cite{hirs95} are nearly identical and close to
the experimental result given by INS Baksan and ITEP+INFN for$\ g_{A}=1.25$.
The same two $M_{2\nu }$ are in agreement with the results of UC-Irvine \cite
{elli91}, ELEGANTS V, LBL and NEMO 3 for $\ g_{A}=1.0$. Further, the NTME $%
M_{2\nu }$ given in the PHFB model, QRPA \cite{suho94} and pseudo-SU3 using
deformed occupation wave functions \cite{hirs95} favor the results of
UC-Irvine \cite{elli91}, ELEGANTS V, LBL, LBL collaboration and ITEP+INFN for%
$\ g_{A}=1.25$. The results of SSDH \cite{simk01} are in agreement with the
experimental half-lives of LBL, LBL collaboration, UC-Irvine \cite{silv97}
and ITEP+INFN. The decay rate $T_{1/2}^{2\nu }$ calculated by Bobyk \textit{%
et al }\cite{boby00}\textit{\ }is in agreement with all the experimental
results due to a large range of values.

The $\left( \beta ^{-}\beta ^{-}\right) _{2\nu }$ decay of $^{104}$Ru$\to
^{104}$Pd for the $0^{+}\to 0^{+}$ transition has not been experimentally
investigated so far. The half-life $T_{1/2}^{2\nu }$ has been calculated
theoretically in QRPA \cite{stau90} and OEM\ \cite{hirs94}. Our predicted $%
T_{1/2}^{2\nu }$ for $g_{A}=1.25$ is approximately 3.7 times larger than
that of calculated in QRPA and smaller than the half-life predicted in OEM
by a factor of approximately 1.3. The predicted $T_{1/2}^{2\nu }$ for $%
g_{A}=1.0$ in the PHFB model is $5.73\times 10^{22}$ yr.

In 1952, Winter had studied the $\left( \beta ^{-}\beta ^{-}\right) _{2\nu }$
decay of $^{110}$Pd isotope for the$\ 0^{+}\to 0^{+}$ transition \cite
{wint52}. The deduced half-life limit was $T_{1/2}^{2\nu }>6.0\times 10^{16}$
yr for $\left( \beta ^{-}\beta ^{-}\right) _{2\nu }$ decay mode and $%
T_{1/2}^{2\nu }>6.0\times 10^{17}$ yr for all modes. Since then, no
experiment has been attempted to study the $\beta ^{-}\beta ^{-}$ decay of $%
^{110}$Pd. The $\left( \beta ^{-}\beta ^{-}\right) _{2\nu }$ decay of $%
^{110} $Pd has been studied theoretically in QRPA \cite{stau90}, OEM \cite
{hirs94}, SRPA \cite{stoi94} and SSDH \cite{civi98,seme00}. The calculated $%
T_{1/2}^{2\nu }$ for $g_{A}=1.25$ in the present PHFB model is $1.74\times
10^{20}$ yr, which is close to that of Semenov \textit{et al }\cite{seme00}
and approximately twice of Civitarese \textit{et al} \cite{civi98}. On the
other hand, the calculated half-life in SRPA is about 7 times larger than
the presently calculated $T_{1/2}^{2\nu }$ for the same $g_{A}$. The
calculated half-life are $1.16\times 10^{19}$ yr and $1.24\times 10^{21}$ yr
in QRPA and OEM respectively. The predicted half-life in the present work
for $g_{A}=1.0$ is $T_{1/2}^{2\nu }=4.24\times 10^{20}$ yr.

In the case of $^{128}$Te nucleus, the presently calculated $M_{2\nu }$ is
close to the NTME extracted from the experiments of Takaoka \textit{et al} 
\cite{taka96}, Manuel \cite{manu91} and Barabash \cite{bara06} for $%
g_{A}=1.0 $ while it is close to those of Lin \textit{et al }\cite{lin88}
and Manuel \cite{manu86} for $g_{A}=1.25$. The NTMEs $M_{2\nu }$ calculated
in SU(4)$_{\sigma \tau }$ \cite{rumy98}, SSDH \cite{seme00} and MCM \cite
{auno96} differ from the presently calculated $M_{2\nu }$ by factor of $%
1.4-1.6$ and are in agreement with the experimental NTMEs due to Lin \textit{%
et al} \cite{lin88} and Manuel \cite{manu86} for $g_{A}=1.0$. The NTME $%
M_{2\nu }$ calculated in SSDH \cite{civi98} is close to the experimental $%
M_{2\nu }$ of Takaoka \textit{et al} \cite{taka96}, Bernatovicz \textit{et al%
} \cite{bern93} and Barabash \cite{bara06} for $g_{A}=1.25$. The presently
calculated $M_{2\nu }$ is smaller than those calculated in QRPA \cite{enge88}
and WCSM \cite{haxt84} by factor of approximately 2.2 and 3.6 respectively
while larger by a factor of approximately 5.5 and 2.5 than the NTMEs
calculated in SRPA \cite{stoi94} and SSDH \cite{civi98} respectively. The
half-life $T_{1/2}^{2\nu }$ calculated in QRPA \cite{stau90} is close to the
experimental $T_{1/2}^{2\nu }$ due to Lin \textit{et al }\cite{lin88} and
Barabash \cite{bara06}. The calculated half-lives $T_{1/2}^{2\nu }$ by
Hirsch \textit{et al} in OEM \cite{hirs94} and Scholten \textit{et al} in
IBM \cite{scho85} are quite small. In the PHFB model, the predicted
half-life $T_{1/2}^{2\nu }$ is $(1.10-2.68)\times 10^{24}$ yr for $%
g_{A}=(1.25-1.0)$.

In the PHFB model, the calculated NTME $M_{2\nu }$ for the $\left( \beta
^{-}\beta ^{-}\right) _{2\nu }$ decay of $^{130}$Te is in agreement with the
NTME extracted from the Milano+INFN experiment \cite{arna03} for $g_{A}=1.0$%
, theoretical $M_{2\nu }$ calculated in the shell model \cite{caur99} and
MCM \cite{auno96}. The calculated NTME $M_{2\nu }$ in SRPA \cite{stoi94} is
in agreement with the experimental $M_{2\nu }$ \cite{lin88,taka96,arna03}
for $g_{A}=1.25$. The presently calculated $M_{2\nu }$ is smaller by a
factor of approximately 1.5 than those calculated in SU(4)$_{\sigma \tau }$ 
\cite{rumy98} and QRPA \cite{enge88}. The $M_{2\nu }$ calculated in RQRPA 
\cite{toiv97} is close to the experimental NTME of Bernatovicz \textit{et al}
\cite{bern93} for $g_{A}=1.25$. The calculated half-lives $T_{1/2}^{2\nu }$
in OEM \cite{hirs94}, IBM \cite{scho85} and WCSM \cite{haxt84} are quite
small while the half-life $T_{1/2}^{2\nu }$ calculated in SU(4)$_{\sigma
\tau }$ \cite{rumy95} is close to the experimentally observed half-lives 
\cite{taka96,lin88,arna03}. The predicted half-life $T_{1/2}^{2\nu }$ in the
PHFB model is $(2.10-5.13)\times 10^{20}$ yr for $g_{A}=(1.25-1.0)$.

In the case of $^{150}$Nd isotope, the calculated $M_{2\nu }$ in the PHFB
model is in agreement with the extracted NTME from NEMO 3 \cite{lala05}
project for $g_{A}=1.25$. The NTME $M_{2\nu }$ calculated in pseudo-SU(3) 
\cite{cast94} is close to the experimental $M_{2\nu }$ of NEMO 3 \cite
{lala05} for $g_{A}=1.0$. The calculated $M_{2\nu }$ in SU(4)$_{\sigma \tau
} $ \cite{rumy98} is larger by a factor of 2.4 than the present value. The
half-life $T_{1/2}^{2\nu }$ calculated in OEM \cite{hirs94} is close to the
experimental result of ITEP+INR project while the calculated $T_{1/2}^{2\nu
} $ in QRPA \cite{stau90} favors the UCI experiment \cite{silv97}. The
calculated half-life $T_{1/2}^{2\nu }$ in the PHFB model is $%
(1.19-2.92)\times 10^{19}$ yr for $g_{A}=(1.25-1.0)$.

In all the cases discussed in table 3, it turns out that the NTMEs $M_{2\nu }
$ calculated with \textit{PQQHH} and \textit{PQQ} effective two-body
effective interactions are very close. Further, it is noticed that although
the calculated spectroscopic properties in the PHFB model for nuclei in the
vibrational limit are not in agreement with the experimental data as
expected, the calculated NTME $M_{2\nu }$ of $^{96}$Zr for example is in
good agreement with the experimentally observed data.

\subsection{Deformation effect}

To understand the role of deformation on the NTMEs $M_{2\nu }$, we
investigate its variation with respect to the change in strength of the 
\textit{QQHH} interaction $\zeta _{qq}$. The results are presented in figure
1. In all cases, it is observed that NTMEs $M_{2\nu }$ remain almost
constant as the strength of $\zeta _{qq}$ is changed from 0.0 to 0.6 except
in case of $^{94}$Zr and $^{100}$Mo isotopes, where $M_{2\nu }$ remains
almost constant for the variation of $\zeta _{qq}$ from 0.0 to 0.8. As $%
\zeta _{qq}$ is further increased up to 1.5, the NTME $M_{2\nu }$ starts
decreasing except a few anomalies. The experimental $M_{2\nu }$ is available
for $\left( \beta ^{-}\beta ^{-}\right) _{2\nu }$ decay of $^{96}$Zr, $%
^{100} $Mo, $^{128,130}$Te and $^{150}$Nd isotopes. It is noteworthy that
the $M_{2\nu }$ tends to be realistic as $\zeta _{qq}$ acquires a physical
value around 1.0. This suggest that the deformations of the HFB intrinsic
states play an important role in reproducing a realistic $M_{2\nu }$.
Further, it is observed in general that there is an anti-correlation between
the NTME $M_{2\nu }$ and the deformation parameter $\beta _{2}$.

\begin{figure}[tbh]
\begin{tabular}{ccc}
\includegraphics [scale=0.53]{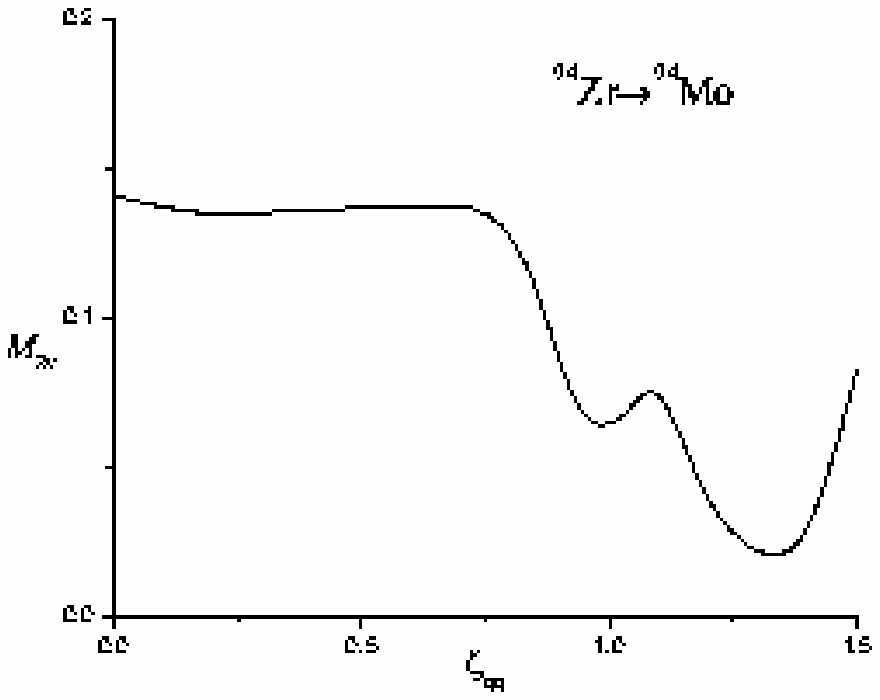} & \includegraphics [scale=0.53]{%
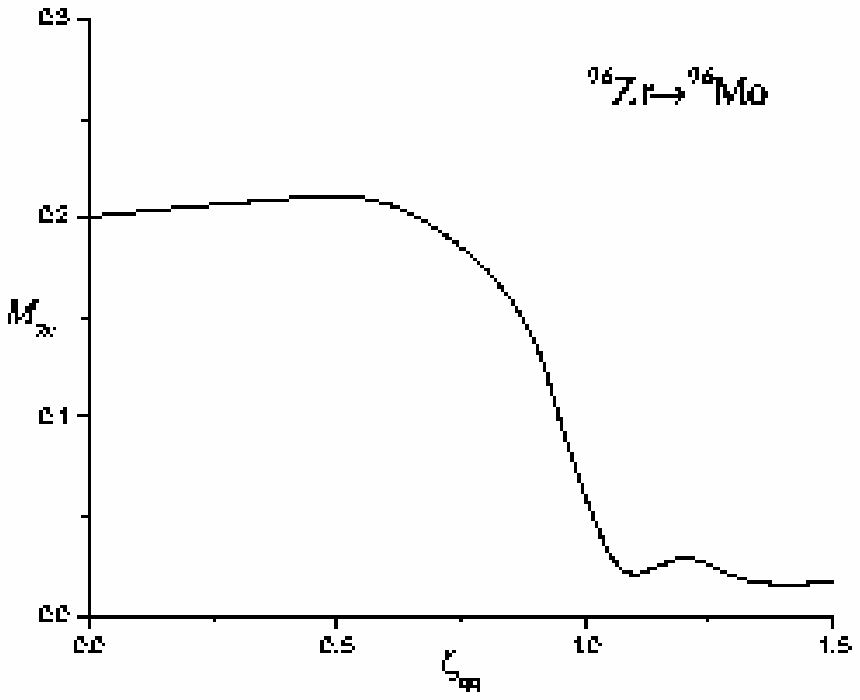} & \includegraphics [scale=0.53]{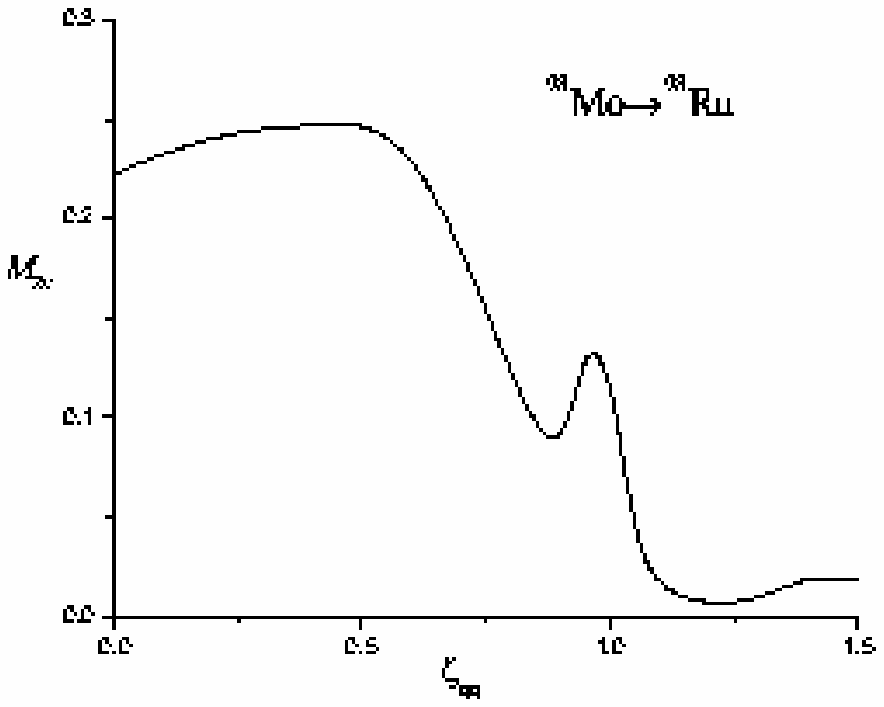} \\ 
\includegraphics [scale=0.53]{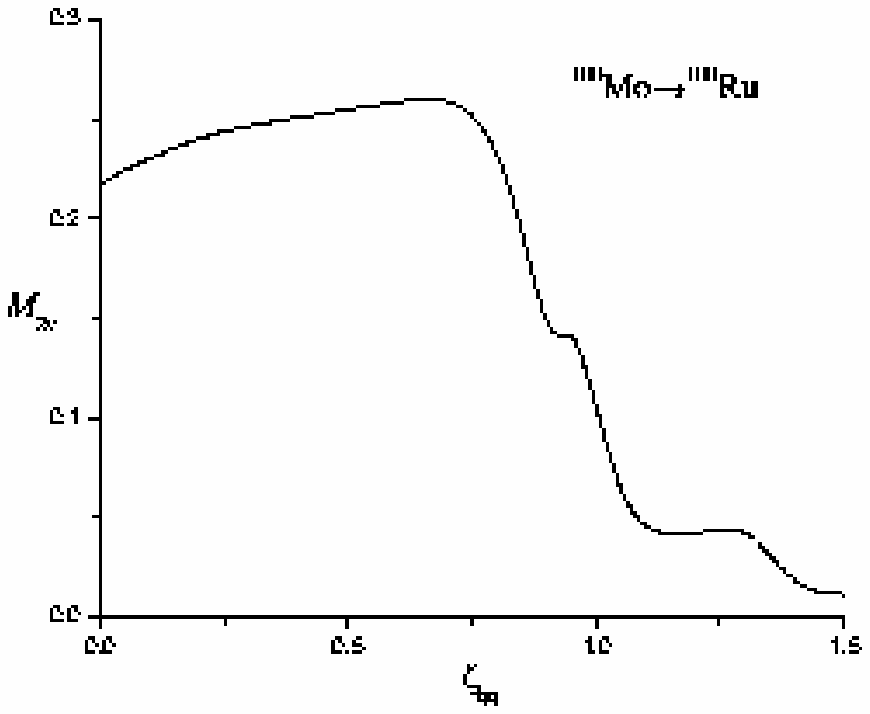} & \includegraphics [scale=0.53]{%
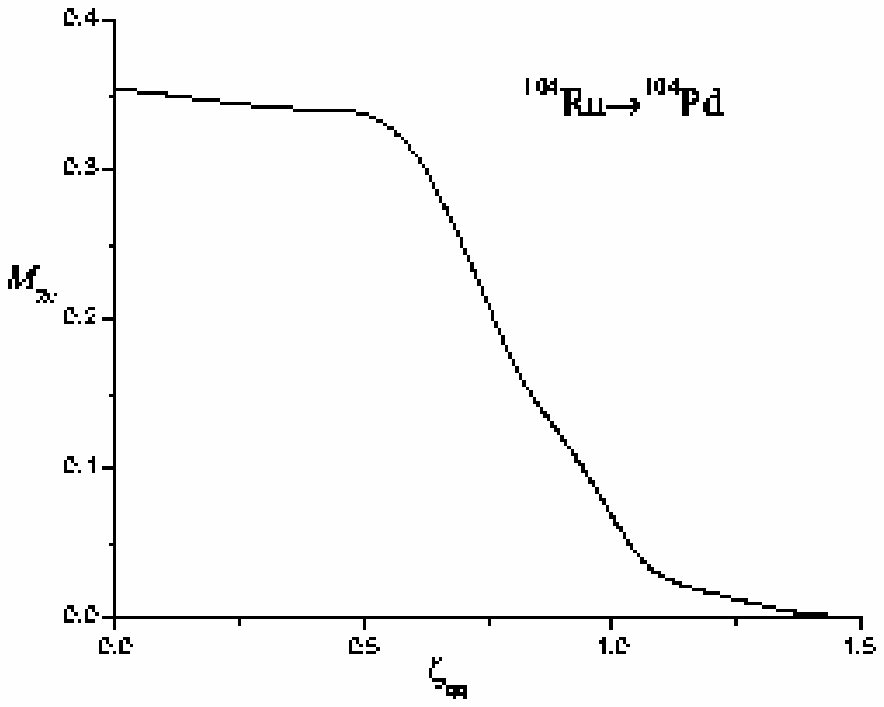} & \includegraphics [scale=0.53]{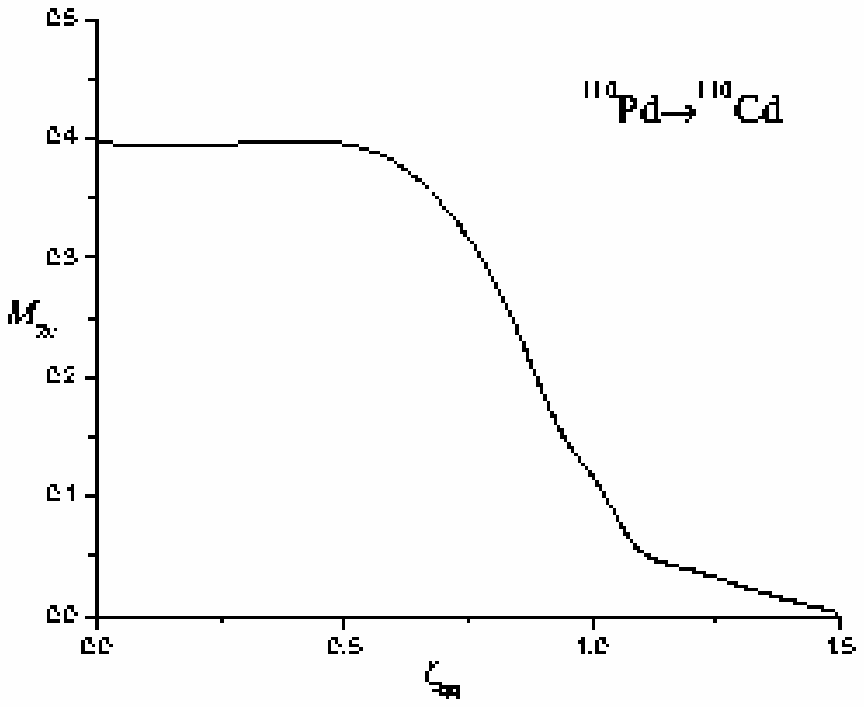} \\ 
\includegraphics [scale=0.53]{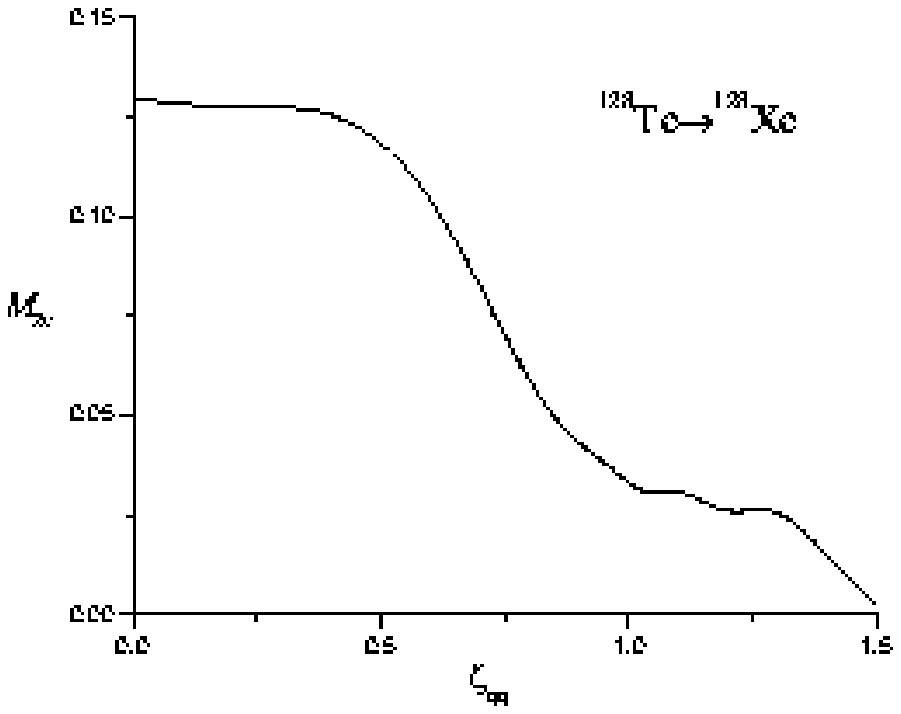} & \includegraphics [scale=0.53]{%
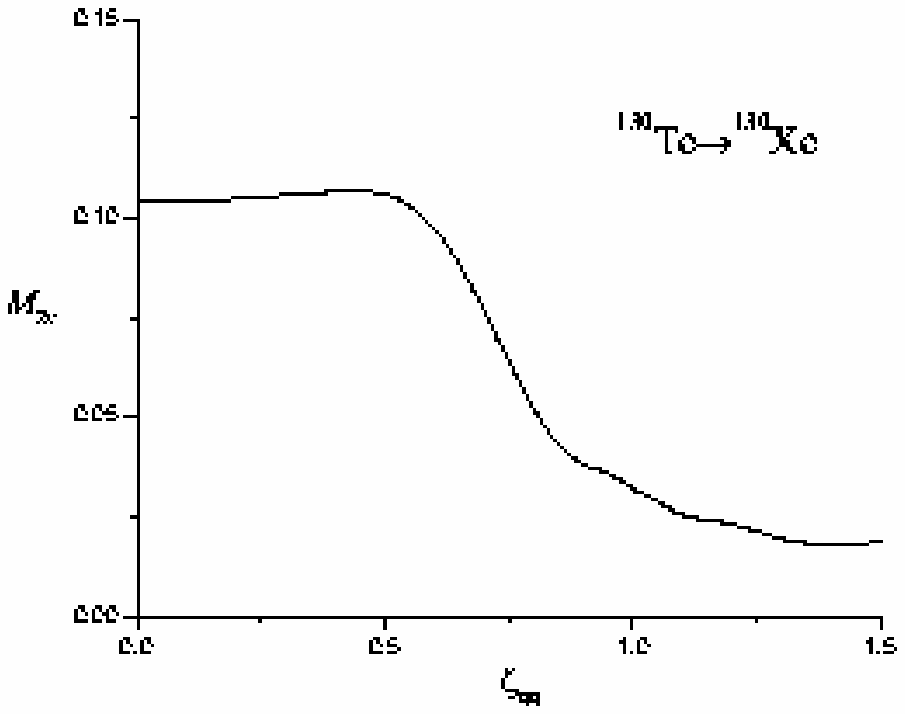} & \includegraphics [scale=0.53]{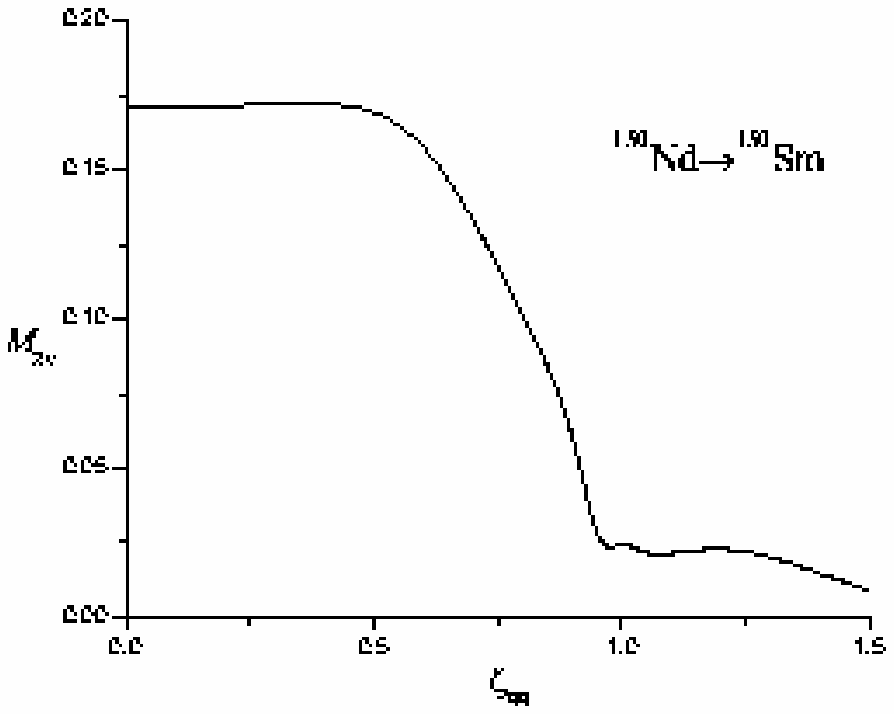}
\end{tabular}
\caption{Dependence of $M_{2\nu }$ on the strength of the $QQHH$ interaction 
$\zeta _{qq}$.}
\end{figure}

To quantify the deformation effect on $M_{2\nu }$, we define a quantity $%
D_{2\nu }$ as the ratio of $M_{2\nu }$ at zero deformation ($\zeta _{qq}=0$)
and full deformation ($\zeta _{qq}=1$). This ratio of deformation effect $%
D_{2\nu }$ is given by 
\begin{equation}
D_{2\nu }=\frac{M_{2\nu }(\zeta _{qq}=0)}{M_{2\nu }(\zeta _{qq}=1)}
\end{equation}
The values of $D_{2\nu }$ for $^{94,96}$Zr, $^{98,100}$Mo, $^{104}$Ru, $%
^{110}$Pd, $^{128,130}$Te and $^{150}$Nd nuclei for $PQQHH(PQQ)$ type of
effective two-body interaction are 2.23(2.29), 3.38(3.70), 1.75(1.86),
2.09(2.33), 5.19(5.47), 3.30(3.14), 3.97(4.26), 3.31(2.89) and 6.46(5.94)
respectively. These values of $D_{2\nu }$ suggest that the $M_{2\nu }$ is
quenched by a factor of approximately 2 to 6.5 in the mass range $A=90-150$
due to deformation effects. Further, the quenching is of almost same
magnitude in the calculation using both type of effective two-body
interaction. We like to mention here that in view of the limitations of the
PHFB\ model as mentioned, the deformation effect for nuclei like $^{94}$Zr
and $^{96}$Zr may be taken as a conservative estmate.

\section{Conclusions}

To summarize, we employ the PHFB model using \textit{PQQHH }type of
effective two-body interaction to construct the yrast band wave functions of 
$\beta \beta $ emitters. The overall agreement between the calculated and
observed yrast spectra as well as electromagnetic properties of the nuclei
suggests that the PHFB wave functions generated by fixing $\chi _{2pn}$ to
reproduce the $E_{2^{+}}$ are quite reliable. Subsequently, we employ the
same wave functions to study the $\left( \beta ^{-}\beta ^{-}\right) _{2\nu
} $ decay of $^{94,96}$Zr, $^{98,100}$Mo, $^{104}$Ru, $^{110}$Pd, $%
^{128,130} $Te and $^{150}$Nd nuclei for the $0^{+}\rightarrow 0^{+}$
transition. The theoretically calculated NTMEs $M_{2\nu }$ are compared with
those extracted from experimentally observed half-lives $T_{1/2}^{2\nu }$.
The agreement between the theoretically calculated results and
experimentally observed data is quite satisfactory. We also examine the
effect of deformation on NTMEs $M_{2\nu }$ by varying the strength of 
\textit{QQHH} part of the effective two-body interaction. It is noticed that
the desired quenching of $M_{2\nu }$ is achieved in the PHFB model through
the subtle interplay of pairing and deformation degrees of freedom. Further,
the results calculated with \textit{PQQ} and \textit{PQQHH} type of
effective two-body interactions are quite similar, which underlines the
stability of our previous results \cite{chan05,sing07}.

Further, a reasonable agreement between the calculated and observed
spectroscopic properties and $M_{2\nu }$ of$\ $ $\left( \beta ^{-}\beta
^{-}\right) _{2\nu }$\ decay for the $0^{+}\rightarrow 0^{+}$ transition of
considered nuclei makes us confident to employ the same PHFB wave functions
to study the $\ $ $\left( \beta ^{-}\beta ^{-}\right) _{0\nu }$\ decay of
the same nuclei, which will be reported soon.

This work was partially supported by DAE-BRNS, India vide sanction No.
2003/37/14/BRNS/669, by Conacyt-Mexico and DGAPA-UNAM.

\pagebreak

\noindent \textbf{Table 1: } Excitation energies $E_{J^{\pi }}$ (MeV) of $%
J^{\pi }=2^{+},$ $4^{+}$ and $6^{+}$ yrast states of $^{94,96}$Zr, $%
^{94,96,98,100}$Mo, $^{98,100,104}$Ru, $^{104,110}$Pd, $^{110}$Cd, $%
^{128,130}$Te, $^{128,130}$ Xe, $^{150}$Nd and $^{150}$Sm nuclei.
\noindent
\begin{tabular}{llllllllll}
\hline\hline
Nucleus &  & \textit{PQQ} & \textit{PQQHH} & Exp.\cite{saka84} & Nucleus & 
& \textit{PQQ} & \textit{PQQHH} & Exp.\cite{saka84} \\ \hline
&  &  &  &  &  &  &  &  &  \\ 
$^{94}$Zr & $\chi _{2pn}$ & 0.02519 & 0.02629 &  & $^{94}$Mo & $\chi _{2pn}$
& 0.02670 & 0.02572 &  \\ 
& $E_{2^{+}}$ & 0.9182 & 0.9165 & 0.9183 &  & $E_{2^{+}}$ & 0.8715 & 0.8713
& 0.871099 \\ 
& $E_{4^{+}}$ & 1.9732 & 1.9657 & 1.4688 &  & $E_{4^{+}}$ & 1.9685 & 1.9682
& 1.573726 \\ 
& $E_{6^{+}}$ & 2.7993 & 2.8087 &  &  & $E_{6^{+}}$ & 3.3136 & 3.3283 & 
2.42337 \\ 
&  &  &  &  &  &  &  &  &  \\ 
$^{96}$Zr & $\chi _{2pn}$ & 0.01717 & 0.01918 &  & $^{96}$Mo & $\chi _{2pn}$
& 0.02557 & 0.02472 &  \\ 
& $E_{2^{+}}$ & 1.7570 & 1.7541 & 1.7507 &  & $E_{2^{+}}$ & 0.7779 & 0.7817
& 0.77821 \\ 
& $E_{4^{+}}$ & 3.5269 & 3.6296 & 3.1202 &  & $E_{4^{+}}$ & 2.0373 & 2.0220
& 1.62815 \\ 
& $E_{6^{+}}$ & 9.7261 & 9.3686 &  &  & $E_{6^{+}}$ & 3.5776 & 3.5300 & 
2.44064 \\ 
&  &  &  &  &  &  &  &  &  \\ 
$^{98}$Mo & $\chi _{2pn}$ & 0.01955 & 0.01955 &  & $^{98}$Ru & $\chi _{2pn}$
& 0.02763 & 0.02649 &  \\ 
& $E_{2^{+}}$ & 0.7892 & 0.7887 & 0.78742 &  & $E_{2^{+}}$ & 0.6513 & 0.6522
& 0.65241 \\ 
& $E_{4^{+}}$ & 1.9522 & 1.9916 & 1.51013 &  & $E_{4^{+}}$ & 1.9430 & 1.9470
& 1.3978 \\ 
& $E_{6^{+}}$ & 3.3098 & 3.4051 & 2.3438 &  & $E_{6^{+}}$ & 3.6548 & 3.6703
& 2.2227 \\ 
&  &  &  &  &  &  &  &  &  \\ 
$^{100}$Mo & $\chi _{2pn}$ & 0.01906 & 0.01876 &  & $^{100}$Ru & $\chi
_{2pn} $ & 0.01838 & 0.01831 &  \\ 
& $E_{2^{+}}$ & 0.5356 & 0.5357 & 0.5355 &  & $E_{2^{+}}$ & 0.5395 & 0.5402
& 0.5396 \\ 
& $E_{4^{+}}$ & 1.4719 & 1.4766 & 1.1359 &  & $E_{4^{+}}$ & 1.5591 & 1.5847
& 1.2265 \\ 
& $E_{6^{+}}$ & 2.6738 & 2.6893 &  &  & $E_{6^{+}}$ & 2.8940 & 2.9629 & 
2.0777 \\ 
&  &  &  &  &  &  &  &  &  \\ 
$^{104}$Ru & $\chi _{2pn}$ & 0.02110 & 0.02053 &  & $^{104}$Pd & $\chi
_{2pn} $ & 0.01486 & 0.01507 &  \\ 
& $E_{2^{+}}$ & 0.3580 & 0.3578 & 0.35799 &  & $E_{2^{+}}$ & 0.5552 & 0.5560
& 0.55579 \\ 
& $E_{4^{+}}$ & 1.1339 & 1.1385 & 0.8885 &  & $E_{4^{+}}$ & 1.5729 & 1.6138
& 1.32359 \\ 
& $E_{6^{+}}$ & 2.2280 & 2.2486 & 1.5563 &  & $E_{6^{+}}$ & 2.8790 & 2.9954
& 2.2498 \\ 
&  &  &  &  &  &  &  &  &  \\ 
$^{110}$Pd & $\chi _{2pn}$ & 0.01417 & 0.01393 &  & $^{110}$Cd & $\chi
_{2pn} $ & 0.01412 & 0.01414 &  \\ 
& $E_{2^{+}}$ & 0.3737 & 0.3738 & 0.3738 &  & $E_{2^{+}}$ & 0.6576 & 0.6585
& 0.6577 \\ 
& $E_{4^{+}}$ & 1.1563 & 1.1583 & 0.9208 &  & $E_{4^{+}}$ & 1.8709 & 1.8921
& 1.5424 \\ 
& $E_{6^{+}}$ & 2.2254 & 2.2359 & 1.5739 &  & $E_{6^{+}}$ & 3.3865 & 3.4728
& 2.4799 \\ 
&  &  &  &  &  &  &  &  &  \\ 
$^{128}$Te & $\chi _{2pn}$ & 0.02715 & 0.02692 &  & $^{128}$Xe & $\chi
_{2pn} $ & 0.0360 & 0.02662 &  \\ 
& $E_{2^{+}}$ & 0.7436 & 0.7435 & 0.7432 &  & $E_{2^{+}}$ & 0.4511 & 0.4420
& 0.4429 \\ 
& $E_{4^{+}}$ & 2.0458 & 2.0130 & 1.4971 &  & $E_{4^{+}}$ & 1.4263 & 1.3444
& 1.0329 \\ 
& $E_{6^{+}}$ & 3.7363 & 3.6647 & 1.8111 &  & $E_{6^{+}}$ & 2.7976 & 2.5404
& 1.7370 \\ 
&  &  &  &  &  &  &  &  &  \\ 
$^{130}$Te & $\chi _{2pn}$ & 0.01801 & 0.01890 &  & $^{130}$Xe & $\chi
_{2pn} $ & 0.02454 & 0.02281 &  \\ 
& $E_{2^{+}}$ & 0.8393 & 0.8395 & 0.8395 &  & $E_{2^{+}}$ & 0.5385 & 0.5384
& 0.5361 \\ 
& $E_{4^{+}}$ & 1.7741 & 1.8085 & 1.6325 &  & $E_{4^{+}}$ & 1.5496 & 1.5268
& 1.2046 \\ 
& $E_{6^{+}}$ & 3.0833 & 3.1283 & 1.8145 &  & $E_{6^{+}}$ & 2.7831 & 2.7400
& 1.9444 \\ 
&  &  &  &  &  &  &  &  &  \\ 
$^{150}$Nd & $\chi _{2pn}$ & 0.02160 & 0.02228 &  & $^{150}$Sm & $\chi
_{2pn} $ & 0.01745 & 0.01730 &  \\ 
& $E_{2^{+}}$ & 0.1307 & 0.1300 & 0.13012 &  & $E_{2^{+}}$ & 0.3328 & 0.3359
& 0.33395 \\ 
& $E_{4^{+}}$ & 0.4320 & 0.4305 & 0.3815 &  & $E_{4^{+}}$ & 1.0156 & 1.0290
& 0.77335 \\ 
& $E_{6^{+}}$ & 0.8960 & 0.8958 &  &  & $E_{6^{+}}$ & 1.9185 & 1.9504 & 
1.27885 \\ 
&  &  &  &  &  &  &  &  &  \\ \hline\hline
\end{tabular}

\pagebreak

\noindent \textbf{Table 2: } Comparison of calculated and experimentally
observed reduced transition probabilities $B(E2$:$0^{+}\rightarrow 2^{+})$,
static quadrupole moments \ $Q(2^{+})$ and $g$-factors $g(2^{+})$ of $%
^{94,96}$Zr, $^{94,96,98,100}$Mo, $^{98,100,104}$Ru, $^{104,110}$Pd, $^{110}$%
Cd, $^{128,130}$Te, $^{128,130}$ Xe, $^{150}$Nd and $^{150}$Sm nuclei. Here $%
B(E2)$ and $Q(2^{+})$ are calculated for effective charge $e_{p}=1+e_{eff}$
and $e_{n}=e_{eff}$. Here $\dagger $ and $\ddagger $ denote the calculation 
with \textit{PQQHH} and \textit{PQQ} type of effective two-body 
interaction respectively.

\noindent
\begin{tabular}{llllllllllllll}
\hline\hline
{\small Nucleus} &  & \multicolumn{4}{c}{$B(E2${\small :}$0^{+}\rightarrow
2^{+})${\small \ (}$e^{2}${\small b}$^{2}${\small )}} &  & 
\multicolumn{4}{c}{$Q(2^{+})${\small \ (}$e${\small b)}} &  & 
\multicolumn{2}{c}{$g(2^{+})${\small \ (nm)}} \\ 
\cline{3-6}\cline{8-11}\cline{13-14}
&  & \multicolumn{3}{l}{\small \ \ \ \ \ \ \ \ \ \ \ \ Theory} & {\small Exp.%
\cite{rama87}} &  & \multicolumn{3}{l}{\small \ \ \ \ \ \ \ \ \ \ Theory} & 
{\small Exp.\cite{ragh89}} &  & {\small Theory} & {\small Exp.\cite{ragh89}}
\\ 
&  & \multicolumn{3}{c}{$e_{eff}$} &  &  & \multicolumn{3}{c}{$e_{eff}$} & 
&  &  &  \\ \cline{3-5}\cline{8-10}
&  & {\small 0.40} & {\small 0.50} & {\small 0.60} &  &  & {\small 0.40} & 
{\small 0.50} & {\small 0.60} &  &  &  &  \\ \hline
&  &  &  &  &  &  &  &  &  &  &  &  &  \\ 
$^{94}${\small Zr} & $\dagger $ & {\small 0.056} & {\small 0.075} & {\small 
\textbf{0.097}} & {\small 0.066}$\pm ${\small 0.014}$^{*}$ & $\dagger $ & 
{\small -0.207} & {\small --0.240} & {\small \textbf{-0.272}} &  & $\dagger $
& {\small 0.112} & {\small -0.329}$\pm ${\small 0.015}$^{a}$ \\ 
& $\ddagger $ & {\small 0.046} & {\small 0.062} & {\small \textbf{0.081}} & 
{\small 0.081}$\pm ${\small 0.017} & $\ddagger $ & {\small -0.168} & {\small %
-0.195} & {\small \textbf{-0.222}} &  & $\ddagger $ & {\small 0.121} & 
{\small -0.26}$\pm ${\small 0.06} \\ 
&  &  &  &  & {\small 0.056}$\pm ${\small 0.014} &  &  &  &  &  &  &  & 
{\small -0.05}$\pm ${\small 0.05} \\ 
&  &  &  &  &  &  &  &  &  &  &  &  &  \\ 
$^{94}${\small Mo} & $\dagger $ & {\small 0.148} & {\small 0.187} & {\small 
\textbf{0.232}} & {\small 0.203}$\pm ${\small 0.004}$^{*}$ & $\dagger $ & 
{\small -0.347} & {\small -0.391} & {\small \textbf{-0.435}} & {\small -0.13}%
$\pm ${\small 0.08} & $\dagger $ & {\small 0.343} &  \\ 
& $\ddagger $ & {\small 0.148} & {\small 0.188} & {\small \textbf{0.232}} & 
{\small 0.230}$\pm ${\small 0.040} & $\ddagger $ & {\small -0.347} & {\small %
-0.391} & {\small \textbf{-0.435}} &  & $\ddagger $ & {\small 0.343} &  \\ 
&  &  &  &  & {\small 0.270}$\pm ${\small 0.035} &  &  &  &  &  &  &  &  \\ 
&  &  &  &  &  &  &  &  &  &  &  &  &  \\ 
$^{96}${\small Zr} & $\dagger $ & {\small 0.046} & {\small \textbf{0.063}} & 
{\small 0.081} & {\small 0.055}$\pm ${\small 0.022}$^{*}$ & $\dagger $ & 
{\small -0.009} & {\small \textbf{-0.011}} & {\small -0.014} &  & $\dagger $
& {\small 0.297} &  \\ 
& $\ddagger $ & {\small 0.044} & {\small \textbf{0.060}} & {\small 0.078} & 
& $\ddagger $ & {\small -0.012} & {\small \textbf{-0.015}} & {\small -0.018}
&  & $\ddagger $ & {\small 0.254} &  \\ 
&  &  &  &  &  &  &  &  &  &  &  &  &  \\ 
$^{96}${\small Mo} & $\dagger $ & {\small 0.250} & {\small \textbf{0.317}} & 
{\small 0.391} & {\small 0.271}$\pm ${\small 0.005}$^{*}$ & $\dagger $ & 
{\small -0.453} & {\small \textbf{-0.509}} & {\small -0.566} & {\small -0.20}%
$\pm ${\small 0.08} & $\dagger $ & {\small 0.535} &  \\ 
& $\ddagger $ & {\small 0.265} & {\small \textbf{0.335}} & {\small 0.413} & 
{\small 0.310}$\pm ${\small 0.047} & $\ddagger $ & {\small -0.466} & {\small 
\textbf{-0.524}} & {\small -0.582} &  & $\ddagger $ & {\small 0.563} &  \\ 
&  &  &  &  & {\small 0.302}$\pm ${\small 0.039} &  &  &  &  &  &  &  &  \\ 
&  &  &  &  &  &  &  &  &  &  &  &  &  \\ 
$^{98}${\small Mo} & $\dagger $ & {\small \textbf{0.246}} & {\small 0.316} & 
{\small 0.395} & {\small 0.267}$\pm ${\small 0.009}$^{*}$ & $\dagger $ & 
{\small \textbf{-0.450}} & {\small -0.510} & {\small -0.570} & {\small -0.26}%
$\pm ${\small 0.09} & $\dagger $ & {\small 0.405} & {\small 0.34}$\pm $%
{\small 0.18} \\ 
& $\ddagger $ & {\small \textbf{0.234}} & {\small 0.302} & {\small 0.378} & 
{\small 0.259}$\pm ${\small 0.010} & $\ddagger $ & {\small \textbf{-0.439}}
& {\small -0.498} & {\small -0.557} &  & $\ddagger $ & {\small 0.376} &  \\ 
&  &  &  &  & {\small 0.260}$\pm ${\small 0.040} &  &  &  &  &  &  &  &  \\ 
&  &  &  &  &  &  &  &  &  &  &  &  &  \\ 
$^{98}${\small Ru} & $\dagger $ & {\small \textbf{0.389}} & {\small 0.488} & 
{\small 0.599} & {\small 0.392}$\pm ${\small 0.012}$^{*}$ & $\dagger $ & 
{\small \textbf{-0.565}} & {\small -0.633} & {\small -0.700} & {\small -0.20}%
$\pm ${\small 0.09} & $\dagger $ & {\small 0.510} & {\small 0.39}$\pm $%
{\small 0.30} \\ 
& $\ddagger $ & {\small \textbf{0.433}} & {\small 0.543} & {\small 0.665} & 
{\small 0.411}$\pm ${\small 0.035} & $\ddagger $ & {\small \textbf{-0.596}}
& {\small -0.667} & {\small -0.739} & {\small -0.03}$\pm ${\small 0.14} & $%
\ddagger $ & {\small 0.528} &  \\ 
&  &  &  &  & {\small 0.475}$\pm ${\small 0.038} &  &  &  &  &  &  &  &  \\ 
&  &  &  &  &  &  &  &  &  &  &  &  &  \\ 
$^{100}${\small Mo} & $\dagger $ & {\small 0.306} & {\small 0.394} & {\small 
\textbf{0.493}} & {\small 0.516}$\pm ${\small 0.010}$^{*}$ & $\dagger $ & 
{\small -0.500} & {\small -0.567} & {\small \textbf{-0.635}} & {\small -0.42}%
$\pm ${\small 0.09} & $\dagger $ & {\small 0.467} & {\small 0.34}$\pm $%
{\small 0.18} \\ 
& $\ddagger $ & {\small 0.320} & {\small 0.412} & {\small \textbf{0.515}} & 
{\small 0.511}$\pm ${\small 0.009} & $\ddagger $ & {\small -0.512} & {\small %
-0.581} & {\small \textbf{-0.650}} & {\small -0.39}$\pm ${\small 0.08} & $%
\ddagger $ & {\small 0.477} &  \\ 
&  &  &  &  & {\small 0.526}$\pm ${\small 0.026} &  &  &  &  &  &  &  &  \\ 
&  &  &  &  &  &  &  &  &  &  &  &  &  \\ 
$^{100}${\small Ru} & $\dagger $ & {\small 0.306} & {\small 0.390} & {\small 
\textbf{0.484}} & {\small 0.490}$\pm ${\small 0.005}$^{*}$ & $\dagger $ & 
{\small -0.501} & {\small -0.565} & {\small \textbf{-0.630}} & {\small -0.54}%
$\pm ${\small 0.07} & $\dagger $ & {\small 0.363} & {\small 0.47}$\pm $%
{\small 0.06} \\ 
& $\ddagger $ & {\small 0.308} & {\small 0.393} & {\small \textbf{0.488}} & 
{\small 0.493}$\pm ${\small 0.003} & $\ddagger $ & {\small -0.503} & {\small %
-0.568} & {\small \textbf{-0.633}} & {\small -0.40}$\pm ${\small 0.12} & $%
\ddagger $ & {\small 0.355} & {\small 0.51}$\pm ${\small 0.07} \\ 
&  &  &  &  & {\small 0.494}$\pm ${\small 0.006} &  &  &  &  & {\small -0.43}%
$\pm ${\small 0.07} &  &  &  \\ 
&  &  &  &  &  &  &  &  &  &  &  &  &  \\ 
$^{104}${\small Ru} & $\dagger $ & {\small 0.558} & {\small 0.714} & {\small 
\textbf{0.890}} & {\small 0.820}$\pm ${\small 0.012}$^{*}$ & $\dagger $ & 
{\small -0.676} & {\small -0.765} & {\small \textbf{-0.854}} & {\small -0.76}%
$\pm ${\small 0.19} & $\dagger $ & {\small 0.345} & {\small 0.41}$\pm $%
{\small 0.05} \\ 
& $\ddagger $ & {\small 0.572} & {\small 0.732} & {\small \textbf{0.912}} & 
{\small 0.93}$\pm ${\small 0.06} & $\ddagger $ & {\small -0.684} & {\small %
-0.774} & {\small \textbf{-0.864}} & {\small -0.70}$\pm ${\small 0.08} & $%
\ddagger $ & {\small 0.339} &  \\ 
&  &  &  &  & {\small 1.04}$\pm ${\small 0.16} &  &  &  &  & {\small -0.66}$%
\pm ${\small 0.05} &  &  &  \\ 
&  &  &  &  &  &  &  &  &  &  &  &  &  \\ \hline
\end{tabular}

\noindent
\begin{tabular}{lllllcllllclcc}
\multicolumn{14}{l}{\small Table 2 continued} \\ \hline
{\small Nucleus} &  & \multicolumn{4}{c}{$B(E2${\small :}$0^{+}\rightarrow
2^{+})${\small \ (}$e^{2}${\small b}$^{2}${\small )}} &  & 
\multicolumn{4}{c}{$Q(2^{+})${\small \ (}$e${\small b)}} &  & 
\multicolumn{2}{c}{$g(2^{+})${\small \ (nm)}} \\ 
\cline{3-6}\cline{8-11}\cline{13-14}
&  & \multicolumn{3}{c}{\small \ Theory} & {\small Exp.\cite{rama87}} &  & 
\multicolumn{3}{c}{\small Theory} & {\small Exp.\cite{ragh89}} &  & {\small %
Theory} & {\small Exp.\cite{ragh89}} \\ 
&  & \multicolumn{3}{c}{$e_{eff}$} & \multicolumn{1}{l}{} &  & 
\multicolumn{3}{c}{$e_{eff}$} & \multicolumn{1}{l}{} &  & \multicolumn{1}{l}{
} & \multicolumn{1}{l}{} \\ \cline{3-5}\cline{8-10}
&  & {\small 0.40} & {\small 0.50} & {\small 0.60} & \multicolumn{1}{l}{} & 
& {\small 0.40} & {\small 0.50} & {\small 0.60} & \multicolumn{1}{l}{} &  & 
\multicolumn{1}{l}{} & \multicolumn{1}{l}{} \\ \hline
&  &  &  &  & \multicolumn{1}{l}{} &  &  &  &  & \multicolumn{1}{l}{} &  & 
\multicolumn{1}{l}{} & \multicolumn{1}{l}{} \\ 
$^{104}${\small Pd} & $\dagger $ & {\small 0.371} & {\small 0.473} & {\small 
\textbf{0.586}} & \multicolumn{1}{l}{{\small 0.535}$\pm ${\small 0.035}$^{*}$%
} & $\dagger $ & {\small -0.550} & {\small -0.621} & {\small \textbf{-0.692}}
& \multicolumn{1}{l}{{\small -0.47}$\pm ${\small 0.10}} & $\dagger $ & 
\multicolumn{1}{l}{\small 0.458} & \multicolumn{1}{l}{{\small 0.46}$\pm $%
{\small 0.04}} \\ 
& $\ddagger $ & {\small 0.361} & {\small 0.460} & {\small \textbf{0.571}} & 
\multicolumn{1}{l}{{\small 0.61}$\pm ${\small 0.09}} & $\ddagger $ & {\small %
-0.543} & {\small -0.613} & {\small \textbf{-0.682}} & \multicolumn{1}{l}{}
& $\ddagger $ & \multicolumn{1}{l}{\small 0.439} & \multicolumn{1}{l}{%
{\small 0.40}$\pm ${\small 0.05}} \\ 
&  &  &  &  & \multicolumn{1}{l}{{\small 0.535}$\pm ${\small 0.035}} &  &  & 
&  & \multicolumn{1}{l}{} &  & \multicolumn{1}{l}{} & \multicolumn{1}{l}{%
{\small 0.38}$\pm ${\small 0.04}} \\ 
&  &  &  &  & \multicolumn{1}{l}{} &  &  &  &  & \multicolumn{1}{l}{} &  & 
\multicolumn{1}{l}{} & \multicolumn{1}{l}{} \\ 
$^{110}${\small Pd} & $\dagger $ & {\small 0.471} & {\small \textbf{0.604}}
& {\small 0.754} & \multicolumn{1}{l}{{\small 0.870}$\pm ${\small 0.040}$%
^{*} $} & $\dagger $ & {\small -0.620} & {\small \textbf{-0.702}} & {\small %
-0.784} & \multicolumn{1}{l}{{\small -0.72}$\pm ${\small 0.14}} & $\dagger $
& \multicolumn{1}{l}{\small 0.489} & \multicolumn{1}{l}{{\small 0.37}$\pm $%
{\small 0.03}} \\ 
& $\ddagger $ & {\small 0.479} & {\small \textbf{0.614}} & {\small 0.766} & 
\multicolumn{1}{l}{{\small 0.780}$\pm ${\small 0.120}} & $\ddagger $ & 
{\small -0.626} & {\small \textbf{-0.708}} & {\small -0.791} & 
\multicolumn{1}{l}{{\small -0.55}$\pm ${\small 0.08}} & $\ddagger $ & 
\multicolumn{1}{l}{\small 0.478} & \multicolumn{1}{l}{{\small 0.35}$\pm $%
{\small 0.03}} \\ 
&  &  &  &  & \multicolumn{1}{l}{{\small 0.820}$\pm ${\small 0.080}} &  &  & 
&  & \multicolumn{1}{l}{{\small -0.47}$\pm ${\small 0.03}} &  & 
\multicolumn{1}{l}{} & \multicolumn{1}{l}{} \\ 
&  &  &  &  & \multicolumn{1}{l}{} &  &  &  &  & \multicolumn{1}{l}{} &  & 
\multicolumn{1}{l}{} & \multicolumn{1}{l}{} \\ 
$^{110}${\small Cd} & $\dagger $ & {\small 0.407} & {\small \textbf{0.522}}
& {\small 0.653} & \multicolumn{1}{l}{{\small 0.450}$\pm ${\small 0.020}$%
^{*} $} & $\dagger $ & {\small -0.575} & {\small \textbf{-0.651}} & {\small %
-0.728} & \multicolumn{1}{l}{{\small -0.40}$\pm ${\small 0.04}} & $\dagger $
& \multicolumn{1}{l}{\small 0.377} & \multicolumn{1}{l}{{\small 0.31}$\pm $%
{\small 0.07}} \\ 
& $\ddagger $ & {\small 0.427} & {\small \textbf{0.548}} & {\small 0.685} & 
\multicolumn{1}{l}{{\small 0.504}$\pm ${\small 0.040}} & $\ddagger $ & 
{\small -0.590} & {\small \textbf{-0.668}} & {\small -0.746} & 
\multicolumn{1}{l}{{\small -0.39}$\pm ${\small 0.05}} & $\ddagger $ & 
\multicolumn{1}{l}{\small 0.358} & \multicolumn{1}{l}{{\small 0.285}$\pm $%
{\small 0.055}} \\ 
&  &  &  &  & \multicolumn{1}{l}{{\small 0.467}$\pm ${\small 0.019}} &  &  & 
&  & \multicolumn{1}{l}{{\small -0.36}$\pm ${\small 0.08}} &  & 
\multicolumn{1}{l}{} & \multicolumn{1}{l}{} \\ 
&  &  &  &  & \multicolumn{1}{l}{} &  &  &  &  & \multicolumn{1}{l}{} &  & 
\multicolumn{1}{l}{} & \multicolumn{1}{l}{} \\ 
$^{128}${\small Te} & $\dagger $ & {\small 0.300} & {\small \textbf{0.384}}
& {\small 0.479} & \multicolumn{1}{l}{{\small 0.383}$\pm ${\small 0.006}$%
^{*} $} & $\dagger $ & {\small -0.498} & {\small \textbf{-0.563}} & {\small %
-0.628} & \multicolumn{1}{l}{{\small -0.14}$\pm ${\small 0.12}} & $\dagger $
& \multicolumn{1}{l}{\small 0.526} & \multicolumn{1}{l}{{\small 0.35}$\pm $%
{\small 0.04}} \\ 
& $\ddagger $ & {\small 0.298} & {\small \textbf{0.381}} & {\small 0.474} & 
\multicolumn{1}{l}{{\small 0.380}$\pm ${\small 0.009}} & $\ddagger $ & 
{\small -0.496} & {\small \textbf{-0.561}} & {\small -0.626} & 
\multicolumn{1}{l}{{\small -0.06}$\pm ${\small 0.05}} & $\ddagger $ & 
\multicolumn{1}{l}{\small 0.514} & \multicolumn{1}{l}{{\small 0.31}$\pm $%
{\small 0.04}} \\ 
&  &  &  &  & \multicolumn{1}{l}{{\small 0.378}$\pm ${\small 0.007}} &  &  & 
&  & \multicolumn{1}{l}{} &  & \multicolumn{1}{l}{} & \multicolumn{1}{l}{}
\\ 
&  &  &  &  & \multicolumn{1}{l}{} &  &  &  &  & \multicolumn{1}{l}{} &  & 
\multicolumn{1}{l}{} & \multicolumn{1}{l}{} \\ 
$^{128}${\small Xe} & $\dagger $ & {\small 0.571} & {\small \textbf{0.729}}
& {\small 0.906} & \multicolumn{1}{l}{{\small 0.750}$\pm ${\small 0.040}$%
^{*} $} & $\dagger $ & {\small -0.685} & {\small \textbf{-0.774}} & {\small %
-0.862} & \multicolumn{1}{l}{} & $\dagger $ & \multicolumn{1}{l}{\small 0.439%
} & \multicolumn{1}{l}{{\small 0.41}$\pm ${\small 0.07}} \\ 
& $\ddagger $ & {\small 0.637} & {\small \textbf{0.819}} & {\small 1.024} & 
\multicolumn{1}{l}{{\small 0.790}$\pm ${\small 0.040}} & $\ddagger $ & 
{\small -0.724} & {\small \textbf{-0.820}} & {\small -0.917} & 
\multicolumn{1}{l}{} & $\ddagger $ & \multicolumn{1}{l}{\small 0.400} & 
\multicolumn{1}{l}{{\small 0.31}$\pm ${\small 0.03}} \\ 
&  &  &  &  & \multicolumn{1}{l}{{\small 0.890}$\pm ${\small 0.230}} &  &  & 
&  & \multicolumn{1}{l}{} &  & \multicolumn{1}{l}{} & \multicolumn{1}{l}{}
\\ 
&  &  &  &  & \multicolumn{1}{l}{} &  &  &  &  & \multicolumn{1}{l}{} &  & 
\multicolumn{1}{l}{} & \multicolumn{1}{l}{} \\ 
$^{130}${\small Te} & $\dagger $ & {\small 0.242} & {\small \textbf{0.304}}
& {\small 0.373} & \multicolumn{1}{l}{{\small 0.295}$\pm ${\small 0.007}$%
^{*} $} & $\dagger $ & {\small -0.448} & {\small \textbf{-0.502}} & {\small %
-0.556} & \multicolumn{1}{l}{{\small -0.15}$\pm ${\small 0.10}} & $\dagger $
& \multicolumn{1}{l}{\small 0.667} & \multicolumn{1}{l}{{\small 0.33}$\pm $%
{\small 0.08}} \\ 
& $\ddagger $ & {\small 0.231} & {\small \textbf{0.289}} & {\small 0.354} & 
\multicolumn{1}{l}{{\small 0.290}$\pm ${\small 0.011}} & $\ddagger $ & 
{\small -0.438} & {\small \textbf{-0.490}} & {\small -0.542} & 
\multicolumn{1}{l}{} & $\ddagger $ & \multicolumn{1}{l}{\small 0.679} & 
\multicolumn{1}{l}{{\small 0.29}$\pm ${\small 0.06}} \\ 
&  &  &  &  & \multicolumn{1}{l}{{\small 0.260}$\pm ${\small 0.050}} &  &  & 
&  & \multicolumn{1}{l}{} &  & \multicolumn{1}{l}{} & \multicolumn{1}{l}{}
\\ 
&  &  &  &  & \multicolumn{1}{l}{} &  &  &  &  & \multicolumn{1}{l}{} &  & 
\multicolumn{1}{l}{} & \multicolumn{1}{l}{} \\ 
$^{130}${\small Xe} & $\dagger $ & {\small 0.478} & {\small \textbf{0.605}}
& {\small 0.746} & \multicolumn{1}{l}{{\small 0.65}$\pm ${\small 0.05}$^{*}$}
& $\dagger $ & {\small -0.627} & {\small \textbf{-0.705}} & {\small -0.783}
& \multicolumn{1}{l}{} & $\dagger $ & \multicolumn{1}{l}{\small 0.478} & 
\multicolumn{1}{l}{{\small 0.38}$\pm ${\small 0.07}} \\ 
& $\ddagger $ & {\small 0.493} & {\small \textbf{0.624}} & {\small 0.769} & 
\multicolumn{1}{l}{{\small 0.631}$\pm ${\small 0.048}} & $\ddagger $ & 
{\small -0.637} & {\small \textbf{-0.716}} & {\small -0.795} & 
\multicolumn{1}{l}{} & $\ddagger $ & \multicolumn{1}{l}{\small 0.463} & 
\multicolumn{1}{l}{{\small 0.31}$\pm ${\small 0.04}} \\ 
&  &  &  &  & \multicolumn{1}{l}{{\small 0.640}$\pm ${\small 0.160}} &  &  & 
&  & \multicolumn{1}{l}{} &  & \multicolumn{1}{l}{} & \multicolumn{1}{l}{}
\\ 
&  &  &  &  & \multicolumn{1}{l}{} &  &  &  &  & \multicolumn{1}{l}{} &  & 
\multicolumn{1}{l}{} & \multicolumn{1}{l}{} \\ 
$^{150}${\small Nd} & $\dagger $ & {\small 2.172} & {\small \textbf{2.632}}
& {\small 3.136} & \multicolumn{1}{l}{{\small 2.760}$\pm ${\small 0.040}$%
^{*} $} & $\dagger $ & {\small -1.335} & {\small \textbf{-1.469}} & {\small %
-1.604} & \multicolumn{1}{l}{{\small -2.00}$\pm ${\small 0.51}} & $\dagger $
& \multicolumn{1}{l}{\small 0.622} & \multicolumn{1}{l}{{\small 0.422}$\pm $%
{\small 0.039}} \\ 
& $\ddagger $ & {\small 2.132} & {\small \textbf{2.580}} & {\small 3.070} & 
\multicolumn{1}{l}{{\small 2.640}$\pm ${\small 0.080}} & $\ddagger $ & 
{\small -1.322} & {\small \textbf{-1.455}} & {\small -1.587} & 
\multicolumn{1}{l}{} & $\ddagger $ & \multicolumn{1}{l}{\small 0.636} & 
\multicolumn{1}{l}{{\small 0.322}$\pm ${\small 0.009}} \\ 
&  &  &  &  & \multicolumn{1}{l}{{\small 2.670}$\pm ${\small 0.100}} &  &  & 
&  & \multicolumn{1}{l}{} &  & \multicolumn{1}{l}{} & \multicolumn{1}{l}{}
\\ 
&  &  &  &  & \multicolumn{1}{l}{} &  &  &  &  & \multicolumn{1}{l}{} &  & 
\multicolumn{1}{l}{} & \multicolumn{1}{l}{} \\ 
$^{150}${\small Sm} & $\dagger $ & {\small 1.741} & {\small \textbf{2.098}}
& {\small 2.488} & \multicolumn{1}{l}{{\small 1.350}$\pm ${\small 0.030}$%
^{*} $} & $\dagger $ & {\small -1.193} & {\small \textbf{-1.310}} & {\small %
-1.426} & \multicolumn{1}{l}{{\small -1.32}$\pm ${\small 0.19}} & $\dagger $
& \multicolumn{1}{l}{\small 0.604} & \multicolumn{1}{l}{{\small 0.385}$\pm $%
{\small 0.027}} \\ 
& $\ddagger $ & {\small 1.707} & {\small \textbf{2.056}} & {\small 2.437} & 
\multicolumn{1}{l}{{\small 1.470}$\pm ${\small 0.090}} & $\ddagger $ & 
{\small -1.182} & {\small \textbf{-1.297}} & {\small -1.412} & 
\multicolumn{1}{l}{{\small -1.25}$\pm ${\small 0.20}} & $\ddagger $ & 
\multicolumn{1}{l}{\small 0.592} & \multicolumn{1}{l}{{\small 0.411}$\pm $%
{\small 0.032}} \\ 
&  &  &  &  & \multicolumn{1}{l}{{\small 1.440}$\pm ${\small 0.150}} &  &  & 
&  & \multicolumn{1}{l}{} &  & \multicolumn{1}{l}{} & \multicolumn{1}{l}{}
\\ 
&  &  &  &  & \multicolumn{1}{l}{} &  &  &  &  & \multicolumn{1}{l}{} &  & 
\multicolumn{1}{l}{} & \multicolumn{1}{l}{} \\ \hline\hline
\end{tabular}
\noindent $^{*}$Average $B(E2)$ values\cite{rama01}, $^{a}$\cite{spei02}

\pagebreak

\noindent \textbf{Table 3: }Experimental half-life $T_{1/2}^{2\nu }$ and
corresponding NTME $M_{2\nu }$ for the $0^{+}\rightarrow 0^{+}$ transition
of $^{94,96}$Zr, $^{98,100}$Mo, $^{104}$Ru, $^{110}$Pd, $^{128,130}$Te and $%
^{150}$Nd nuclei along with theoretically calculated NTME $M_{2\nu }$ and
half-life $T_{1/2}^{2\nu }$ in different nuclear models. The numbers
corresponding to (a) and (b) are calculated for $g_{A}=1.25$ and 1.0
respectively. ``gch.'' denotes the geochemical experiment and ``*'' denotes
the present calculation.

\noindent 
\begin{tabular}{llllcllllcl}
\hline\hline
\multicolumn{6}{c}{\small Experiment} & \multicolumn{5}{c}{\small Theory} \\ 
{\small Nuclei} & {\small Project} & {\small Ref.} & $T_{1/2}^{2\nu }$ &  & $%
\left| M_{2\nu }\right| $ & {\small Model} & {\small Ref.} & $\left| M_{2\nu
}\right| $ &  & $T_{1/2}^{2\nu }$ \\ \hline
&  &  &  &  &  &  &  &  &  &  \\ 
$^{94}${\small Zr} & {\small NEMO 2} & {\small \cite{arno99}} & $>${\small %
1.1}$\times ${\small 10}$^{-5}$ & {\small (a)} & $<${\small 62.815} & 
{\small PHFB}$^{*}$ & \textit{PQQHH} & {\small 0.063} & {\small (a)} & 
{\small 10.86} \\ 
{\small (10}$^{22}${\small \ yr)} &  &  &  & {\small (b)} & $<${\small 98.148%
} &  &  &  & {\small (b)} & {\small 26.52} \\ 
&  &  &  &  &  &  & \textit{PQQ} & {\small 0.076} & {\small (a)} & {\small %
7.51} \\ 
&  &  &  &  &  &  &  &  & {\small (b)} & {\small 18.34} \\ 
&  &  &  &  &  & {\small SRQRPA} & {\small \cite{boby00}} &  &  & {\small %
3.08-659} \\ 
&  &  &  &  &  & {\small OEM} & {\small \cite{hirs94}} &  &  & {\small 168}
\\ 
&  &  &  &  &  & {\small QRPA} & {\small \cite{stau90}} &  &  & {\small 6.93}
\\ 
&  &  &  &  &  &  &  &  &  &  \\ 
$^{96}${\small Zr} & {\small NEMO 3} & {\small \cite{lala05}} & {\small 2.0}$%
\pm ${\small 0.3}$\pm ${\small 0.2} & {\small (a)} & {\small 0.051}$%
_{-0.0054}^{+0.0079}$ & {\small PHFB}$^{*}$ & \textit{PQQHH} & {\small 0.059}
& {\small (a)} & {\small 1.47} \\ 
{\small (10}$^{19}${\small \ yr)} &  &  &  & {\small (b)} & {\small 0.080}$%
_{-0.0084}^{+0.0123}$ &  &  &  & {\small (b)} & {\small 3.59} \\ 
& {\small gch.} & {\small \cite{wies01}} & {\small 0.94}$\pm ${\small 0.32}
& {\small (a)} & {\small 0.074}$_{-0.0101}^{+0.0172}$ &  & \textit{PQQ} & 
{\small 0.058} & {\small (a)} & {\small 1.56} \\ 
&  &  &  & {\small (b)} & {\small 0.116}$_{-0.0158}^{+0.0269}$ &  &  &  & 
{\small (b)} & {\small 3.80} \\ 
& {\small NEMO 2} & {\small \cite{bara98}} & {\small 2.0}$_{-0.5}^{+0.9}\pm $%
{\small 0.5} & {\small (a)} & {\small 0.051}$_{-0.012}^{+0.021}$ & {\small %
SRQRPA} & {\small \cite{boby00}} &  &  & {\small 0.452-61} \\ 
&  &  &  & {\small (b)} & {\small 0.080}$_{-0.019}^{+0.033}$ & {\small SU(4)}%
$_{\sigma \tau }$ & {\small \cite{rumy98}} & {\small 0.0678} & {\small (a)}
& {\small 1.13} \\ 
& {\small gch}$.$ & {\small \cite{kawa93}} & {\small 3.9}$\pm ${\small 0.9}
& {\small (a)} & {\small 0.036}$_{-0.0036}^{+0.0051}$ &  &  &  & {\small (b)}
& {\small 2.76} \\ 
&  &  &  & {\small (b)} & {\small 0.057}$_{-0.0056}^{+0.0080}$ & {\small %
RQRPA}$^{\dagger }$ & {\small \cite{toiv97}} &  &  & {\small 4.4} \\ 
& {\small Average} & {\small \cite{elli02}} & {\small 1.4}$_{-0.5}^{+3.5}$ & 
{\small (a)} & {\small 0.061}$_{-0.0283}^{+0.0151}$ & {\small RQRPA}$%
^{\ddagger }$ & {\small \cite{toiv97}} &  &  & {\small 4.2} \\ 
& {\small Value} &  &  & {\small (b)} & {\small 0.095}$_{-0.0443}^{+0.0235}$
& {\small QRPA} & {\small \cite{bara96}} & {\small 0.12-0.31} & {\small (a)}
& {\small 0.054-0.36} \\ 
& {\small Recommended} & {\small \cite{bara06}} & {\small 2.0}$\pm 0.3$ & 
{\small (a)} & {\small 0.051}$_{-0.0034}^{+0.0043}$ &  &  &  & {\small (b)}
& {\small 0.13-0.88} \\ 
& {\small Value} &  &  & {\small (b)} & {\small 0.078}$_{-0.0054}^{+0.0067}$
& {\small SRPA} & {\small \cite{stoi95}} & {\small 0.022} & {\small (a)} & 
{\small 10.72} \\ 
&  &  &  &  &  &  &  &  & {\small (b)} & {\small 26.18} \\ 
&  &  &  &  &  & {\small OEM} & {\small \cite{hirs94}} &  &  & {\small 20.2}
\\ 
&  &  &  &  &  & {\small QRPA} & {\small \cite{stau90}} &  &  & {\small 1.08}
\\ 
&  &  &  &  &  & {\small QRPA} & {\small \cite{enge88}} & {\small 0.124} & 
{\small (a)} & {\small 0.34} \\ 
&  &  &  &  &  &  &  &  & {\small (b)} & {\small 0.82} \\ 
&  &  &  &  &  &  &  &  &  &  \\ 
$^{98}${\small Mo} &  &  &  &  &  & {\small PHFB}$^{*}$ & \textit{PQQHH} & 
{\small 0.127} & {\small (a)} & {\small 6.36} \\ 
{\small (10}$^{29}${\small \ yr)} &  &  &  &  &  &  &  &  & {\small (b)} & 
{\small 15.52} \\ 
&  &  &  &  &  &  & \textit{PQQ} & {\small 0.130} & {\small (a)} & {\small %
6.09} \\ 
&  &  &  &  &  &  &  &  & {\small (b)} & {\small 14.87} \\ 
&  &  &  &  &  & {\small SRQRPA} & {\small \cite{boby00}} &  &  & {\small %
4.06-15.2} \\ 
&  &  &  &  &  & {\small OEM} & {\small \cite{hirs94}} &  &  & {\small 61.6}
\\ 
&  &  &  &  &  & {\small QRPA} & {\small \cite{stau90}} &  &  & {\small 29.6}
\\ 
&  &  &  &  &  &  &  &  &  &  \\ 
$^{100}${\small Mo} & {\small NEMO 3} & {\small \cite{arno05}} & {\small 7.11%
}$\pm ${\small 0.02}$\pm ${\small 0.54} & {\small (a)} & {\small 0.122}$%
_{-0.0045}^{+0.0051}$ & {\small PHFB}$^{*}$ & \textit{PQQHH} & {\small 0.104}
& {\small (a)} & {\small 9.71} \\ 
{\small (10}$^{18}${\small \ yr)} &  &  &  & {\small (b)} & {\small 0.191}$%
_{-0.0071}^{+0.0080}$ &  &  &  & {\small (b)} & {\small 23.72} \\ 
& {\small NEMO 3} & {\small \cite{lala05}} & {\small 7.72}$\pm ${\small 0.02}%
$\pm ${\small 0.54} & {\small (a)} & {\small 0.117}$_{-0.0040}^{+0.0045}$ & 
& \textit{PQQ} & {\small 0.104} & {\small (a)} & {\small 9.79} \\ 
&  &  &  & {\small (b)} & {\small 0.183}$_{-0.0063}^{+0.0070}$ &  &  &  & 
{\small (b)} & {\small 23.90} \\ 
& {\small gch.} & {\small \cite{hida04}} & {\small 2.1}$\pm ${\small 0.3} & 
{\small (a)} & {\small 0.225}$_{-0.0145}^{+0.0180}$ & {\small SSDH} & 
{\small \cite{simk01}} &  & {\small (a)} & {\small 7.15-8.97} \\ 
&  &  &  & {\small (b)} & {\small 0.351}$_{-0.0227}^{+0.0281}$ & {\small %
SRQRPA} & {\small \cite{boby00}} &  &  & {\small 5.04-16800} \\ 
&  &  &  &  &  &  &  &  &  &  \\ \hline
\end{tabular}

\noindent \noindent 
\begin{tabular}{llllcllllcl}
\multicolumn{11}{l}{\small Table 3 continued} \\ \hline
\multicolumn{6}{c}{\small Experiment} & \multicolumn{5}{c}{\small Theory} \\ 
{\small Nuclei} & {\small Project} & {\small Ref.} & $T_{1/2}^{2\nu }$ &  & $%
\left| M_{2\nu }\right| $ & {\small Model} & {\small Ref.} & $\left| M_{2\nu
}\right| $ &  & $T_{1/2}^{2\nu }$ \\ \hline
&  &  &  &  &  &  &  &  &  &  \\ 
& {\small ITEP+INFN} & {\small \cite{ashi01}} & {\small 7.2}$\pm ${\small 0.9%
}$\pm ${\small 1.8} & {\small (a)} & {\small 0.121}$_{-0.0179}^{+0.0321}$ & 
{\small SU(4)}$_{\sigma \tau }$ & {\small \cite{rumy98}} & {\small 0.1606} & 
{\small (a)} & {\small 4.11} \\ 
&  &  &  & {\small (b)} & {\small 0.190}$_{-0.0279}^{+0.0502}$ &  &  &  & 
{\small (b)} & {\small 10.03} \\ 
& {\small UC Irvine} & {\small \cite{silv97}} & {\small 6.82}$%
_{-0.53}^{+0.38}\pm 0.68$ & {\small (a)} & {\small 0.125}$%
_{-0.0087}^{+0.0128}$ & {\small SSDH} & {\small \cite{civi98}} & {\small 0.18%
} & {\small (a)} & {\small 3.27} \\ 
&  &  &  & {\small (b)} & {\small 0.195}$_{-0.0136}^{+0.0200}$ &  &  &  & 
{\small (b)} & {\small 7.99} \\ 
& {\small LBL+MHC+} & {\small \cite{alst97}} & {\small 7.6}$_{-1.4}^{+2.2}$
& {\small (a)} & {\small 0.118}$_{-0.0141}^{+0.0127}$ & {\small SRPA} & 
{\small \cite{stoi95}} & {\small 0.059} & {\small (a)} & {\small 30.45} \\ 
& {\small UNM+INEL} &  &  & {\small (b)} & {\small 0.185}$%
_{-0.0220}^{+0.0198}$ &  &  &  & {\small (b)} & {\small 74.34} \\ 
& {\small NEMO} & {\small \cite{dass95}} & {\small 9.5}$\pm ${\small 0.4}$%
\pm ${\small 0.9} & {\small (a)} & {\small 0.106}$_{-0.007}^{+0.008}$ & 
{\small pSU(3)}$^{+}$ & {\small \cite{hirs95}} & {\small 0.152} & {\small (a)%
} & {\small 4.59} \\ 
&  &  &  & {\small (b)} & {\small 0.165}$_{-0.010}^{+0.013}$ &  &  &  & 
{\small (b)} & {\small 11.2} \\ 
& {\small LBL} & {\small \cite{garc93}} & {\small 9.7}$\pm 4.9$ & {\small (a)%
} & {\small 0.105}$_{-0.0193}^{+0.0441}$ & {\small pSU(3)}$^{++}$ & {\small 
\cite{hirs95}} & {\small 0.108} & {\small (a)} & {\small 9.09} \\ 
&  &  &  & {\small (b)} & {\small 0.163}$_{-0.0302}^{+0.0689}$ &  &  &  & 
{\small (b)} & {\small 22.19} \\ 
& {\small ELEGANTS V} & {\small \cite{ejir91}} & {\small 11.5}$%
_{-2.0}^{+3.0} $ & {\small (a)} & {\small 0.096}$_{-0.0105}^{+0.0096}$ & 
{\small OEM} & {\small \cite{hirs94}} &  &  & {\small 35.8} \\ 
&  &  &  & {\small (b)} & {\small 0.150}$_{-0.0164}^{+0.0150}$ & {\small QRPA%
} & {\small \cite{suho94}} & {\small 0.101} & {\small (a)} & {\small 10.39}
\\ 
& {\small UC Irvine} & {\small \cite{elli91}} & {\small 11.6}$_{-0.8}^{+3.4}$
& {\small (a)} & {\small 0.096}$_{-0.0115}^{+0.0035}$ &  &  &  & {\small (b)}
& {\small 25.37} \\ 
&  &  &  & {\small (b)} & {\small 0.149}$_{-0.0180}^{+0.0054}$ & {\small QRPA%
} & {\small \cite{grif92}} & {\small 0.256} & {\small (a)} & {\small 1.62}
\\ 
& {\small INS Baksan} & {\small \cite{vasi90}} & {\small 3.3}$_{-1.0}^{+2.0}$
& {\small (a)} & {\small 0.179}$_{-0.0378}^{+0.0355}$ &  &  &  & {\small (b)}
& {\small 3.95} \\ 
&  &  &  & {\small (b)} & {\small 0.280}$_{-0.0591}^{+0.0554}$ & {\small QRPA%
} & {\small \cite{stau90}} &  &  & {\small 1.13} \\ 
& {\small Average} & {\small \cite{elli02}} & {\small 8.0}$\pm 0.6$ & 
{\small (a)} & {\small 0.115}$_{-0.0047}^{+0.0054}$ & {\small QRPA} & 
{\small \cite{enge88}} & {\small 0.211} & {\small (a)} & {\small 2.38} \\ 
& {\small Value} &  &  & {\small (b)} & {\small 0.180}$_{-0.0074}^{+0.0084}$
&  &  &  & {\small (b)} & {\small 5.81} \\ 
& {\small Average} & {\small \cite{bara06}} & {\small 7.1}$\pm 0.4$ & 
{\small (a)} & {\small 0.122}$_{-0.0033}^{+0.0036}$ &  &  &  &  &  \\ 
& {\small Value} &  &  & {\small (b)} & {\small 0.191}$_{-0.0052}^{+0.0056}$
&  &  &  &  &  \\ 
&  &  &  &  &  &  &  &  &  &  \\ 
$^{104}${\small Ru} &  &  &  &  &  & {\small PHFB}$^{*}$ & \textit{PQQHH} & 
{\small 0.068} & {\small (a)} & {\small 2.35} \\ 
{\small (10}$^{22}${\small \ yr)} &  &  &  &  &  &  &  &  & {\small (b)} & 
{\small 5.73} \\ 
&  &  &  &  &  &  & \textit{PQQ} & {\small 0.068} & {\small (a)} & {\small %
2.35} \\ 
&  &  &  &  &  &  &  &  & {\small (b)} & {\small 5.73} \\ 
&  &  &  &  &  & {\small OEM} & {\small \cite{hirs94}} &  &  & {\small 3.09}
\\ 
&  &  &  &  &  & {\small QRPA} & {\small \cite{stau90}} &  &  & {\small 0.629%
} \\ 
&  &  &  &  &  &  &  &  &  &  \\ 
$^{110}${\small Pd} &  & {\small \cite{wint52}} & $>${\small 6.0}$\times $%
{\small 10}$^{-4}$ & {\small (a)} & $<${\small 6.468} & {\small PHFB}$^{*}$
& \textit{PQQHH} & {\small 0.120} & {\small (a)} & {\small 1.74} \\ 
{\small (10}$^{20}${\small \ yr)} &  &  &  & {\small (b)} & $<${\small 10.106%
} &  &  &  & {\small (b)} & {\small 4.24} \\ 
&  &  &  &  &  &  & \textit{PQQ} & {\small 0.133} & {\small (a)} & {\small %
1.41} \\ 
&  &  &  &  &  &  &  &  & {\small (b)} & {\small 3.44} \\ 
&  &  &  &  &  & {\small SSDH} & {\small \cite{seme00}} &  &  & {\small 1.6}
\\ 
&  &  &  &  &  & {\small SSDH} & {\small \cite{civi98}} & {\small 0.19} & 
{\small (a)} & {\small 0.7} \\ 
&  &  &  &  &  &  &  &  & {\small (b)} & {\small 1.70} \\ 
&  &  &  &  &  & {\small SRPA} & {\small \cite{stoi94}} & {\small 0.046} & 
{\small (a)} & {\small 11.86} \\ 
&  &  &  &  &  &  &  &  & {\small (b)} & {\small 28.96} \\ 
&  &  &  &  &  & {\small OEM} & {\small \cite{hirs94}} &  &  & {\small 12.4}
\\ 
&  &  &  &  &  & {\small QRPA} & {\small \cite{stau90}} &  &  & {\small 0.116%
} \\ 
&  &  &  &  &  &  &  &  &  &  \\ \hline
\end{tabular}

\noindent 
\begin{tabular}{lllllllllll}
\multicolumn{11}{l}{\small Table 3 continued} \\ \hline
\multicolumn{6}{c}{\small Experiment} & \multicolumn{5}{c}{\small Theory} \\ 
{\small Nuclei} & {\small Project} & {\small Ref.} & $T_{1/2}^{2\nu }$ &  & $%
\left| M_{2\nu }\right| $ & {\small Model} & {\small Ref.} & $\left| M_{2\nu
}\right| $ &  & $T_{1/2}^{2\nu }$ \\ \hline
&  &  &  & \multicolumn{1}{c}{} &  &  &  &  & \multicolumn{1}{c}{} &  \\ 
$^{128}${\small Te} & {\small gch.} & {\small \cite{taka96}} & {\small (2.2}$%
\pm ${\small 0.3)} & {\small (a)} & {\small 0.023}$_{-0.0014}^{+.00018}$ & 
{\small PHFB}$^{*}$ & \textit{PQQHH} & {\small 0.033} & {\small (a)} & 
{\small 1.10} \\ 
{\small (10}$^{24}${\small \ yr)} &  &  &  & {\small (b)} & {\small 0.036}$%
_{-0.0022}^{+0.0027}$ &  &  &  & {\small (b)} & {\small 2.68} \\ 
& {\small gch.} & {\small \cite{bern93}} & {\small 7.7}$\pm ${\small 0.4} & 
{\small (a)} & {\small 0.012}$_{-0.0003}^{+0.0003}$ &  & \textit{PQQ} & 
{\small 0.033} & {\small (a)} & {\small 1.05} \\ 
&  &  &  & {\small (b)} & {\small 0.019}$_{-0.0005}^{+0.0005}$ &  &  &  & 
{\small (b)} & {\small 2.55} \\ 
& {\small gch.} & {\small \cite{manu91}} & {\small 2.0} & {\small (a)} & 
{\small 0.024} & {\small SSDH} & {\small \cite{seme00}} & {\small 0.048} & 
{\small (a)} & {\small 0.51} \\ 
&  &  &  & {\small (b)} & {\small 0.038} &  &  &  & {\small (b)} & {\small %
1.29} \\ 
& {\small gch.} & {\small \cite{lin88}} & {\small 1.8}$\pm ${\small 0.7} & 
\multicolumn{1}{c}{\small (a)} & {\small 0.026}$_{-0.0039}^{+0.0071}$ & 
{\small SM} & {\small \cite{caur99}} &  & \multicolumn{1}{c}{} & {\small 0.5}
\\ 
&  &  &  & \multicolumn{1}{c}{\small (b)} & {\small 0.040}$%
_{-0.0060}^{+0.0111}$ & {\small SU(4)}$_{\sigma \tau }$ & {\small \cite
{rumy98}} & {\small 0.053} & \multicolumn{1}{c}{\small (a)} & {\small 0.42}
\\ 
& {\small gch.} & {\small \cite{manu86}} & {\small 1.4}$\pm ${\small 0.4} & 
\multicolumn{1}{c}{\small (a)} & {\small 0.029}$_{-0.0034}^{+0.0053}$ &  & 
&  & \multicolumn{1}{c}{\small (b)} & {\small 1.06} \\ 
&  &  &  & \multicolumn{1}{c}{\small (b)} & {\small 0.045}$%
_{-0.0053}^{+0.0083}$ & {\small SSDH} & {\small \cite{civi98}} & {\small %
0.013} & \multicolumn{1}{c}{\small (a)} & {\small 6.98} \\ 
& {\small Average} & {\small \cite{elli02}} & {\small 7.2}$\pm 0.3$ & 
\multicolumn{1}{c}{\small (a)} & {\small 0.013}$_{-0.0003}^{+0.0003}$ &  & 
&  & \multicolumn{1}{c}{\small (b)} & {\small 17.65} \\ 
& {\small Value} &  &  & \multicolumn{1}{c}{\small (b)} & {\small 0.020}$%
_{-0.0004}^{+0.0004}$ & {\small MCM} & {\small \cite{auno96}} & {\small 0.046%
} & \multicolumn{1}{c}{\small (a)} & {\small 0.56} \\ 
& {\small Recommended} & {\small \cite{bara06}} & {\small 2.5}$\pm ${\small %
0.4} & \multicolumn{1}{c}{\small (a)} & {\small 0.022}$_{-0.0012}^{+0.0014}$
&  &  &  & \multicolumn{1}{c}{\small (b)} & {\small 1.41} \\ 
& {\small Value} &  &  & \multicolumn{1}{c}{\small (b)} & {\small 0.034}$%
_{-0.0019}^{+0.0022}$ & {\small SRPA} & {\small \cite{stoi94}} & {\small %
0.006} & \multicolumn{1}{c}{\small (a)} & {\small 32.78} \\ 
&  &  &  & \multicolumn{1}{c}{} &  &  &  &  & \multicolumn{1}{c}{\small (b)}
& {\small 82.87} \\ 
&  &  &  & \multicolumn{1}{c}{} &  & {\small OEM} & {\small \cite{hirs94}} & 
& \multicolumn{1}{c}{} & {\small 0.21} \\ 
&  &  &  & \multicolumn{1}{c}{} &  & {\small QRPA} & {\small \cite{stau90}}
&  & \multicolumn{1}{c}{} & {\small 2.63} \\ 
&  &  &  & \multicolumn{1}{c}{} &  & {\small QRPA} & {\small \cite{enge88}}
& {\small 0.074} & \multicolumn{1}{c}{\small (a)} & {\small 0.22} \\ 
&  &  &  & \multicolumn{1}{c}{} &  &  &  &  & \multicolumn{1}{c}{\small (b)}
& {\small 0.54} \\ 
&  &  &  & \multicolumn{1}{c}{} &  & {\small IBM} & {\small \cite{scho85}} & 
& \multicolumn{1}{c}{} & {\small 0.09} \\ 
&  &  &  & \multicolumn{1}{c}{} &  & {\small WCSM} & {\small \cite{haxt84}}
& {\small 0.120} & \multicolumn{1}{c}{\small (a)} & {\small 0.08} \\ 
&  &  &  & \multicolumn{1}{c}{} &  &  &  &  & \multicolumn{1}{c}{\small (b)}
& {\small 0.21} \\ 
&  &  &  & \multicolumn{1}{c}{} &  &  &  &  & \multicolumn{1}{c}{} &  \\ 
$^{130}${\small Te} & {\small Milano+INFN} & {\small \cite{arna03}} & 
{\small 6.1}$\pm ${\small 1.4}$_{-3.5}^{+2.9}$ & \multicolumn{1}{c}{\small %
(a)} & {\small 0.018}$_{-0.0043}^{+0.0232}$ & {\small PHFB}$^{*}$ & \textit{%
PQQHH} & {\small 0.031} & \multicolumn{1}{c}{\small (a)} & {\small 2.10} \\ 
{\small (10}$^{20}${\small \ yr)} &  &  &  & \multicolumn{1}{c}{\small (b)}
& {\small 0.029}$_{-0.0068}^{+0.0362}$ &  &  &  & \multicolumn{1}{c}{\small %
(b)} & {\small 5.13} \\ 
& {\small gch.} & {\small \cite{taka96}} & {\small 7.9}$\pm ${\small 1.0} & 
{\small (a)} & {\small 0.016}$_{-0.0009}^{+0.0011}$ &  & \textit{PQQ} & 
{\small 0.042} & {\small (a)} & {\small 1.16} \\ 
&  &  &  & {\small (b)} & {\small 0.025}$_{-0.0015}^{+0.0018}$ &  &  &  & 
{\small (b)} & {\small 2.82} \\ 
& {\small gch.} & {\small \cite{bern93}} & {\small 27.0}$\pm ${\small 1.0} & 
\multicolumn{1}{c}{\small (a)} & {\small 0.009}$_{-0.0002}^{+0.0002}$ & 
{\small SM} & {\small \cite{caur99}} & {\small 0.030} & \multicolumn{1}{c}%
{\small (a)} & {\small 2.3} \\ 
&  &  &  & \multicolumn{1}{c}{\small (b)} & {\small 0.014}$%
_{-0.0002}^{+0.0003}$ &  &  &  & \multicolumn{1}{c}{\small (b)} & {\small %
5.84} \\ 
& {\small gch} & {\small \cite{lin88}} & {\small 7.5}$\pm ${\small 0.3} & 
\multicolumn{1}{c}{\small (a)} & {\small 0.017}$_{-0.0003}^{+0.0003}$ & 
{\small SU(4)}$_{\sigma \tau }$ & {\small \cite{rumy98}} & {\small 0.0468} & 
\multicolumn{1}{c}{\small (a)} & {\small 0.95} \\ 
&  &  &  & \multicolumn{1}{c}{\small (b)} & {\small 0.026}$%
_{-0.0005}^{+0.0005}$ &  &  &  & \multicolumn{1}{c}{\small (b)} & {\small %
2.32} \\ 
& {\small Average} & {\small \cite{elli02}} & {\small 27}$\pm ${\small 1.0}
& \multicolumn{1}{c}{\small (a)} & {\small 0.009}$_{-0.0002}^{+0.0002}$ & 
{\small RQRPA}$^{\dagger }$ & {\small \cite{toiv97}} & {\small 0.009} & 
\multicolumn{1}{c}{\small (a)} & {\small 25.68} \\ 
& {\small .Value} &  &  & \multicolumn{1}{c}{\small (b)} & {\small 0.014}$%
_{-0.0002}^{+0.0003}$ &  &  &  & \multicolumn{1}{c}{\small (b)} & {\small %
62.70} \\ 
& {\small Recommended} & {\small \cite{bara06}} & {\small 9.0}$\pm ${\small %
1.0} & \multicolumn{1}{c}{\small (a)} & {\small 0.015}$_{-0.0008}^{+0.0009}$
& {\small RQRPA}$^{\ddagger }$ & {\small \cite{toiv97}} & {\small 0.009} & 
\multicolumn{1}{c}{\small (a)} & {\small 25.68} \\ 
& {\small Value} &  &  & \multicolumn{1}{c}{\small (b)} & {\small 0.024}$%
_{-0.0012}^{+0.0014}$ &  &  &  & \multicolumn{1}{c}{\small (b)} & {\small %
62.70} \\ 
&  &  &  & \multicolumn{1}{c}{} &  & {\small MCM} & {\small \cite{auno96}} & 
{\small 0.028} & \multicolumn{1}{c}{\small (a)} & {\small 2.65} \\ 
&  &  &  & \multicolumn{1}{c}{} &  &  &  &  & \multicolumn{1}{c}{\small (b)}
& {\small 6.48} \\ 
&  &  &  & \multicolumn{1}{c}{} &  & {\small SU(4)}$_{\sigma \tau }$ & 
{\small \cite{rumy95}} &  & \multicolumn{1}{c}{} & {\small 7.0} \\ 
&  &  &  & \multicolumn{1}{c}{} &  & {\small SRPA} & {\small \cite{stoi94}}
& {\small 0.016} & \multicolumn{1}{c}{\small (a)} & {\small 8.12} \\ 
&  &  &  & \multicolumn{1}{c}{} &  &  &  &  & \multicolumn{1}{c}{\small (b)}
& {\small 19.84} \\ 
&  &  &  & \multicolumn{1}{c}{} &  & {\small OEM} & {\small \cite{hirs94}} & 
& \multicolumn{1}{c}{} & {\small 0.79} \\ 
&  &  &  & \multicolumn{1}{c}{} &  & {\small QRPA} & {\small \cite{stau90}}
&  & \multicolumn{1}{c}{} & {\small 18.4} \\ 
&  &  &  & \multicolumn{1}{c}{} &  & {\small QRPA} & {\small \cite{enge88}}
& {\small 0.049} & \multicolumn{1}{c}{\small (a)} & {\small 0.87} \\ 
&  &  &  & \multicolumn{1}{c}{} &  &  &  &  & \multicolumn{1}{c}{\small (b)}
& {\small 2.12} \\ 
&  &  &  & \multicolumn{1}{c}{} &  & {\small IBM} & {\small \cite{scho85}} & 
& \multicolumn{1}{c}{} & {\small 0.17} \\ 
&  &  &  & \multicolumn{1}{c}{} &  & {\small WCSM} & {\small \cite{haxt84}}
& {\small 0.114} & \multicolumn{1}{c}{\small (a)} & {\small 0.16} \\ 
&  &  &  & \multicolumn{1}{c}{} &  &  &  &  & \multicolumn{1}{c}{\small (b)}
& {\small 0.40} \\ 
&  &  &  &  &  &  &  &  &  &  \\ \hline
\end{tabular}

\noindent 
\begin{tabular}{lllllllllll}
\multicolumn{11}{l}{\small Table 3 continued} \\ \hline
\multicolumn{6}{c}{\small Experiment} & \multicolumn{5}{c}{\small Theory} \\ 
{\small Nuclei} & {\small Project} & {\small Ref.} & $T_{1/2}^{2\nu }$ &  & $%
\left| M_{2\nu }\right| $ & {\small Model} & {\small Ref.} & $\left| M_{2\nu
}\right| $ &  & $T_{1/2}^{2\nu }$ \\ \hline
&  &  &  &  &  &  &  &  &  &  \\ 
$^{150}${\small Nd} & {\small NEMO 3} & {\small \cite{lala05}} & {\small 9.7}%
$\pm ${\small 0.7}$\pm ${\small 1.0} & \multicolumn{1}{c}{\small (a)} & 
{\small 0.029}$_{-0.0023}^{+0.0030}$ & {\small PHFB}$^{*}$ & \textit{PQQHH}
& {\small 0.027} & \multicolumn{1}{c}{\small (a)} & {\small 11.95} \\ 
{\small (10}$^{18}${\small \ yr)} &  &  &  & \multicolumn{1}{c}{\small (b)}
& {\small 0.046}$_{-0.0036}^{+0.0047}$ &  &  &  & \multicolumn{1}{c}{\small %
(b)} & {\small 29.17} \\ 
&  &  &  &  &  &  & \textit{PQQ} & {\small 0.033} & {\small (a)} & {\small %
7.89} \\ 
&  &  &  &  &  &  &  &  & {\small (b)} & {\small 19.27} \\ 
& {\small UCI} & {\small \cite{silv97}} & {\small 6.75}$_{-0.42}^{+0.37}\pm
0.68$ & \multicolumn{1}{c}{\small (a)} & {\small 0.035}$_{-0.0025}^{+0.0033}$
& {\small SU(4)}$_{\sigma \tau }$ & {\small \cite{rumy98}} & {\small 0.0642}
& \multicolumn{1}{c}{\small (a)} & {\small 2.04} \\ 
&  &  &  & \multicolumn{1}{c}{\small (b)} & {\small 0.055}$%
_{-0.0038}^{+.00051}$ &  &  &  & \multicolumn{1}{c}{\small (b)} & {\small %
4.98} \\ 
& {\small ITEP +INR} & {\small \cite{arte95}} & {\small 18.8}$%
_{-3.9}^{+6.6}\pm 1.9$ & \multicolumn{1}{c}{\small (a)} & {\small 0.021}$%
_{-0.0036}^{+0.0043}$ & {\small pSU(3)} & {\small \cite{cast94}} & {\small %
0.055} & \multicolumn{1}{c}{\small (a)} & {\small 2.78} \\ 
&  &  &  & \multicolumn{1}{c}{\small (b)} & {\small 0.033}$%
_{-0.0056}^{+0.0067}$ &  &  &  & \multicolumn{1}{c}{\small (b)} & {\small %
6.79} \\ 
& {\small ITEP +INR} & {\small \cite{arte93}} & {\small 17}$_{-5.0}^{+10}\pm 
${\small 3.5} & \multicolumn{1}{c}{\small (a)} & {\small 0.022}$%
_{-0.0056}^{+0.0092}$ & {\small OEM} & {\small \cite{hirs94}} &  & 
\multicolumn{1}{c}{} & {\small 16.6} \\ 
&  &  &  & \multicolumn{1}{c}{\small (b)} & {\small 0.035}$%
_{-0.0088}^{+0.0144}$ & {\small QRPA} & {\small \cite{stau90}} &  & 
\multicolumn{1}{c}{} & {\small 7.37} \\ 
& {\small Average} & {\small \cite{elli02}} & {\small 7.0}$_{-0.3}^{+11.8}$
& \multicolumn{1}{c}{\small (a)} & {\small 0.035}$_{-0.0135}^{+0.0008}$ &  & 
&  & \multicolumn{1}{c}{} &  \\ 
& {\small Value} &  &  & \multicolumn{1}{c}{\small (b)} & {\small 0.054}$%
_{-0.0211}^{+0.0012}$ &  &  &  & \multicolumn{1}{c}{} &  \\ 
& {\small Average} & {\small \cite{bara06}} & {\small 7.8}$\pm ${\small 0.7}
& \multicolumn{1}{c}{\small (a)} & {\small 0.033}$_{-0.0014}^{+0.0016}$ &  & 
&  & \multicolumn{1}{c}{} &  \\ 
& {\small Value} &  &  & \multicolumn{1}{c}{\small (b)} & {\small 0.051}$%
_{-0.0022}^{+0.0025}$ &  &  &  & \multicolumn{1}{c}{} &  \\ \hline\hline
\end{tabular}

\noindent $^{\dagger }$AWS basis; $^{\ddagger }$WS basis; $^{+}$Spherical
occupation wave functions; $^{++}$Deformed occupation wave functions

\end{document}